\documentstyle[epsf]{elsart}
\begin{document}
\begin{frontmatter}
\title{Electroinduced two-nucleon knockout and correlations in nuclei}
\author{J. Ryckebusch \thanksref{email}}, 
\author{V. Van der Sluys},
\author{K. Heyde}, 
\author{H. Holvoet}, 
\author{W. Van Nespen} 
and  
\author{M. Waroquier} 
\address{ Department of Subatomic and Radiation Physics
\protect\\University of Gent
\protect\\ Proeftuinstraat 86, B-9000 Gent, Belgium}
\author{M. Vanderhaeghen}
\address{CEA DAPNIA-SPhN, C.E. Saclay, France}
\thanks[email]{E-mail : jan.ryckebusch@rug.ac.be} 
\begin{abstract}
We present a model to calculate cross sections for electroinduced
two-nucleon emission from finite nuclei.  Short-range correlations in
the wave functions and meson-exchange contributions to the
photoabsorption process are implemented.  Effects of the short-range
correlations are studied with the aid of a perturbation expansion
method with various choices of the Jastrow correlation function.  The
model is used to investigate the relative importance of the different
reaction mechanisms contributing to the A(e,e$'$pn) and A(e,e$'$pp)
process.   Representative examples for the target nuclei $^{12}$C and
$^{16}$O and for kinematical conditions accessible with contemporary
high-duty cycle electron accelerators are presented.
A procedure is outlined to calculate the two-nucleon knockout contribution
to the semi-exclusive (e,e$'$p) cross section.  
Using this technique we investigate in how far semi-exclusive
(e,e$'$p) reactions can be used to detect high-momentum components in
the nuclear spectral function.\protect 
\end{abstract}
\end{frontmatter}
\section{Introduction}
The strong short-range and tensor component in realistic nucleon-nucleon
interactions induces correlations in the nuclear many-body wave
functions that cannot be accounted for in an independent-particle
approximation as for example adopted in the Hartree-Fock (HF) model. 
Over the years, various models that aim at going beyound the
independent-particle model (IPM) wave functions, including effects of 
short-range and
tensor correlations, have been proposed \cite{mahan,pieper,co,della,gua96}.
Despite the strongness of the
tensor and short-range force it turns out to be a very challenging task
to determine the proper measurable quantities that can verify the realistic
character of those models.  Cross sections for pion double-charge 
exchange reactions $_Z$A($\pi ^+$,$\pi ^-$)$_{Z+2}$A have been 
shown to exhibit
some sensitivity to dynamical short-range correlations (SRC)
 but cannot yet distinguish between
the different model predictions \cite{wirz89,johnson}.  

Better conditions to study the effects of correlations are
predicted for reactions induced by leptonic probes.  In contrast to
pion-induced reactions, the data are not contaminated by effects
related to initial-state interactions and the whole nuclear volume,
including the interior, is sampled.   It is believed that with the
advent of a new
generation of high-duty cycle electron facilities in the intermediate energy
range ($\epsilon = 0.5 - 4$~GeV) good conditions have been created 
to explore the dynamics of
ground-state correlations with unprecedented precision.

A key function in the study of ground-state correlations in nuclei is
the single-particle spectral function $P(\vec{k},E)$ that gives the
probability to remove a nucleon with momentum $\vec{k}$ and find the
residual A-1 system at an energy
$E$.  
The most dramatic effects of the short-range and tensor correlations
on $P(\vec{k},E)$ are predicted to occur at high momentum and energy
\cite{ciofi96,mutter}.
One should realize, however, that these parts of the spectral function
belong to the smaller probability 
components in the nuclear wave functions.  As a consequence,
signals of ground-state correlations are likely to produce
relatively small cross sections.  This puts heavy constraints on the
experimental requirements when performing measurements 
that aim at probing those correlations.
Moreover, when exploring
 the small components in the nuclear wave
functions one should start worrying about the role of competing
mechanisms that could blur their effects 
in the actual cross sections.  In this respect it is worth mentioning
that the frequently quoted relation between the (e,e$'$p) cross
section and the single-particle spectral function   
is based on rather severe assumptions regarding the reaction
mechanism.  Apart from the neglect of any sort of final-state
interaction effect, the initial photoabsorption mechanism is assumed
to be dominated by one-body operators.   Even under quasi-elastic
conditions, when this assumption is believed to have good chances to
match reality, there are indications for many-body effects
contributing to the (e,e$'$p) cross sections \cite{wei94,Wei89,vemec}.

Already in the late fifties it was suggested that two-nucleon knockout
reactions might shed some light on short-range correlations in nuclei.
The underlying idea is that a photon hitting a strongly correlated
nucleon pair will induce two nucleons to escape from the target
\cite{Got58}. 
Two-nucleon knockout reactions have been extensively studied using 
real photons.  In these investigations, data have been taken in
a wide photon energy range (E$_{\gamma} < $ 1 GeV), covering both the
proton-proton and proton-neutron knockout channel.  Rather than
revealing information about ground-state correlations, the gross
features of the $(\gamma,pp)$ and $(\gamma,pn)$ data could be
interpreted in terms of meson-exchange and pion degrees of freedom
\cite{cross,grab96,lamp96,harty96}. 
Data for electroinduced two-nucleon knockout reactions
are rather scarce.  Pioneering experiments have been performed at
NIKHEF-K \cite{zondervan,leon}. More data will be collected at MAMI and
TJNAF in the near future. It is hoped that the additional longitudinal
degree of freedom will create optimum circumstances to reveal
signatures of ground-state correlations.

One of the principal aims of this paper is to provide a framework in
which models for ground-state correlations can be confronted with
cross sections for electroinduced two-nucleon knockout reactions.  In
our developments we leave room for reaction mechanisms that can feed
the two-nucleon knockout channel and cannot be directly associated
with ground-state correlations.  The major competing reaction
mechanism consists of processes in which the photon is absorbed on
two-, three-, ... nucleon currents. Just as the ground-state
correlation effects these multi-body currents are a natural
manifestation of the many-body dynamics of the nuclear system.  For
the present purposes we will restrict ourselves to two-body currents
related to one-pion exchange. The framework in which those currents
are treated, however, is rather general and can be extended to include
the effects of heavier meson exchange. Apart from the exclusive 
two-nucleon
knockout channel, we address also the (e,e$'$p) reaction.  We
concentrate on the high missing-energy region where the effects of
ground-state correlations are predicted to manifest themselves.  The
exclusive nature of the electroproton reaction at high missing
energies cannot be guaranteed.  In our theoretical considerations, the
starting point will be that the single-particle spectral function is
not a directly measurable quantity.  It will be argued that
two-nucleon knockout represents an important fraction of the continuum
(e,e$'$p) strength and that the (e,e$'$p)
can be considered as a possible source of information about
many-body dynamics in the nuclear system.

The outline of this paper is as follows.  The specific
ingredients of our model are introduced in Sec.~\ref{formalism}.         
This includes a discussion of a practical way of calculating the
exclusive (e,e$'$NN) (Sec.~\ref{sec:eenn}) and semi-inclusive (e,e$'$N)
(Sec.~\ref{sec:semi}) cross section.  The model assumptions with
respect to the meson-exchange and isobaric currents are outlined in
Sec.~\ref{sec:mec} and those related to ground-state correlations in
Sec.~\ref{sec:src}. The results of the (e,e$'$pp), (e,e$'$pn) and
semi-inclusive (e,e$'$p) calculations are discussed in
Sec.~\ref{sec:resu}. 
    
\section{Formalism}
\label{formalism}
\subsection{The (e,e$'$NN) cross section}
\label{sec:eenn}
\begin{figure}
\centering
\epsfysize=10.cm
\epsffile{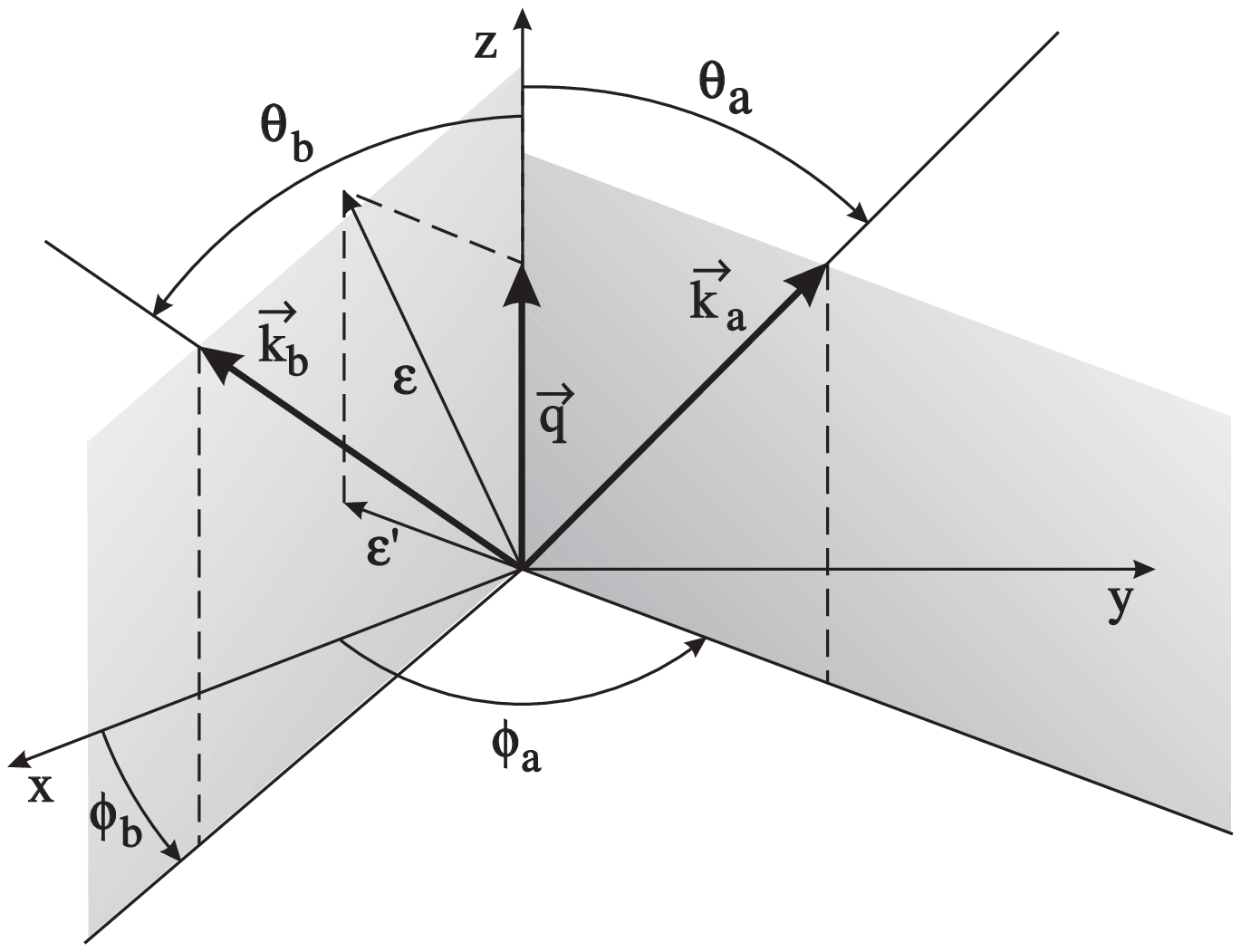}
\caption{Adopted conventions for the angular variables of an
A(e,e$'$N$_a$N$_b$) reaction.  The electron scattering plane is the xz plane.}
\label{fig:kine}
\end{figure}

In the plane-wave approximation for the incoming and scattered electron
waves the cross section for triple coincidence reactions of the type
(e,e$'$N$_a$N$_b$) can be written as :
\begin{eqnarray}
& & {d^5 \sigma \over dE_b d \Omega _b d \Omega _a d \epsilon ' d \Omega
_{\epsilon '}} (e,e'N_aN_b) = 
{1 \over 4 (2\pi)^8 } k_a k_b E_a E_b f_{rec} \sigma_{M}
\nonumber \\ 
\times
& & \biggl[ v_T W_T(\theta _a,\phi _a,\theta _b,\phi _b)
+ v_C W_L(\theta _a,\phi _a,\theta _b,\phi _b) 
\nonumber \\ 
& &
+
v_I W_{LT}(\theta _a,\phi _a,\theta _b,\phi _b) +
v_S W_{TT} (\theta _a,\phi _a,\theta _b,\phi _b) \biggr] \;,
\label{eq:eepnn}
\end{eqnarray}
where the Mott cross section $\sigma _M$ is given by :
\begin{equation}
\sigma _M = \frac {e^2 cos^2 {\theta _e \over 2}} {4 \epsilon ^2 
sin^4 {\theta _e \over 2}} \; .
\end{equation}
For the cross section (\ref{eq:eepnn}) we have considered the
situation in which the residual $A-2$ nucleus is created at a fixed
excitation energy  $E_x$, which is expressed 
relative to its ground-state energy.  As a
consequence, the integration over the energy of one of the escaping
particles has been performed.
The functions $v$ contain all the electron kinematics and read :
\begin{eqnarray}
v_T &=& tg^2 \frac{\theta_e}{2} -
\frac{1}{2}\left(\frac{q_{\mu}q^{\mu}}{\vec{q}^2}\right)  \\
v_C &=& \left( \frac{q_{\mu}}{\vec{q}} \right)^4
\\ 
v_I&=&\frac{q_{\mu}q^{\mu}}
{\sqrt{2} \mid \vec{q} \mid ^3} (\epsilon + \epsilon ')
tg \frac{\theta_e}{2}
\\ 
v_S&=&\frac{q_{\mu}q^{\mu}}{2 \vec{q} ^2} \; .
\end{eqnarray}
The recoil factor in the above cross section reads :
\begin{equation}
f_{rec} = \frac {1} {1+ \frac {E_a} {E_{A-2}} \left(
1 - \frac {q cos \theta _a} {k_a} + \frac {k_b cos \theta_{ab}}
{k_a} \right)} \;,
\end{equation}
where $\theta _{ab}$ is the angle between the directions of the ejected
nucleons. 
The structure functions $W$ are defined in terms of the electromagnetic
transition operators between the ground state of the target nucleus and
the final state and are functions of the polar ($\theta$) and azimuthal
($\phi$) angle of the two escaping particles.  Our choice for the
reference frame and the notation conventions
regarding the kinematical variables are summarized in
Fig.~\ref{fig:kine}. In deriving the above
expression we have not considered any polarization condition for the
final products and the electrons.  
Accordingly, the structure functions $W$ involve a sum
over the spin projections $(m_{s_{a}},m_{s_{b}}$ and $M_R$) of the final
products, including two escaping nucleons and a residual nucleus created
in a specified state $\mid \Psi _f ^{(A-2)} (E_x,J_R M_R) >$ :
\begin{eqnarray}
W_L(\theta _a,\phi _a,\theta _b,\phi _b) & = &\sum_{m_{s_{a}},m_{s_{b}},M_R}
\left(m_F^{fi}(\lambda=0)\right)^*\left(m_F^{fi}(\lambda=0)\right)
\label{wl}
 \\
W_T(\theta _a,\phi _a,\theta _b,\phi _b) & = & 
\sum_{m_{s_{a}},m_{s_{b}},M_R} \left[
\left( m_F^{fi}(\lambda=+1)\right)^*\left(m_F^{fi}(\lambda=+1)\right)
\right. \nonumber \\  
& & +
\left. \left(m_F^{fi}(\lambda=-1)\right)^*\left(m_F^{fi}(\lambda=-1)
\right) \right]
\label{wt}
\\ 
W_{LT}(\theta _a,\phi _a,\theta _b,\phi _b) & = & 2 Re 
\Biggl[ 
\sum_{m_{s_{a}},m_{s_{b}},M_R}\left[
\left(m_F^{fi}(\lambda=0)\right)^*\left(m_F^{fi}(\lambda=-1)
\right) \right.
\nonumber \\ 
& &  \left. 
- \left(m_F^{fi}(\lambda=0)\right)^*\left(m_F^{fi}(\lambda=+1)\right)
\right] \Biggr]
\\ 
W_{TT}(\theta _a,\phi _a,\theta _b,\phi _b) & = & 2 Re \Biggl[
\sum_{m_{s_{a}},m_{s_{b}},M_R}
\left(m_F^{fi}(\lambda=-1)\right)^*
\nonumber \\ 
& & \times \left(m_F^{fi}(\lambda=+1)\right)
\Biggr] \;,
\label{wtt}
\end{eqnarray}
with
\begin{equation}
m_F^{fi} (\lambda = \pm 1)
 =  \left< \Psi_{f}^{(A-2)}(E_x,J_R M_R);
{\vec k}_a m_{s_{a}};{\vec k}_b m_{s_{b}}
\mid J_{\lambda} ({\vec q}) \mid \Psi _0 \right> \\
\label{mfia}
\end{equation}
\begin{equation} 
m_F^{fi} (\lambda = 0)
 = \left< \Psi_{f}^{(A-2)}(E_x,J_R M_R);{\vec k}_a m_{s_{a}};{\vec k}_b m_{s_{b}}
\mid \rho ({\vec q}) \mid \Psi _0 \right> \; .
\label{mfib}
\end{equation}
Here, $J_{\lambda = \pm 1}$ stands for the transverse components
of the nuclear current density, $\rho$ is the nuclear charge density
and $ \mid \Psi _0 >$ is the ground-state wave function of the target
system. 
It goes without saying that computing the above matrix elements with a
final state characterized by two scattering states can be very involving
when accounting for the full complexity of the final-state interaction (FSI)
and that several assumptions have to be made in order to keep the
calculations feasible.  In our model calculations we have put special
emphasis on the orthogonality condition between the initial and final
states thus avoiding spurious contributions entering the matrix elements. This
is of particular importance in view of the fact that on several
occasions we will make integrations over relatively large parts
of the available phase space for
the final products.   In two-nucleon knockout from finite nuclei
the final-state
interaction involves apart from mutual interactions between the
escaping nucleons also distorting effects from the residual $A-2$
system.  Another aspect of the FSI that deserves further attention is
the role of multi-step processes.   An obvious way to take these into
account is a coupled-channel calculation.  In the two-nucleon emission
case, however, lots of channels are expected to contribute which makes a
coupled-channel calculation very challenging.  In the absence of a
consistent and handleable theory to describe the full complexity of the
FSI, we adopt the view that rather than accounting for only part of the
FSI it is maybe better to work in a direct knockout model.  Or put in
other words, we assume that the two nucleons involved in the
photoabsorption mechanism will escape from the target nucleus without
being subject to inelastic collisions with the core.  This is the
basic assumption of the so-called ``spectator approximation''. 
In the shell-model
picture, direct emission of two nucleons will leave
the residual nucleus in a
two-hole (2h) state relative to the ground state
of the target nucleus.  Even in a direct
reaction model, the residual $A-2$ nucleons will distort the wave
function of both escaping nucleons.  In our model calculations this FSI
effect is implemented along the lines explained in Ref.~\cite{jannpa}.  
There it is shown that a proper A-body wave function with two
asymptotically free nucleons and
a residual nucleus generated in a 2h state can be reached by
performing a partial wave expansion in terms of the 
two-hole-two-particle eigenstates of a mean-field one-body Hamiltonian.  
For the
sake of completeness we mention the obtained expression for the
required type of final wave function :
\begin{eqnarray}
\mid \Psi_f > & \equiv & \mid \Psi_{f}^{(A-2)}(E_x,J_R M_R);{\vec{k}}_a
m_{s_{a}} ; {\vec{k}}_b m_{s_{b}} > \nonumber \\
& = & \sum_{lm_ljm}\sum_{l'm_{l'}j'm'} \sum_{JMJ_{1}M_{1}} 
(4 \pi)^2 i^{l+l'}  
\frac {\pi} {2 M_N \sqrt{k_a k_b}} 
\nonumber 
e^{i(\delta_l+\sigma_l 
+\delta_{l'}+\sigma_{l'})}
\\ & & 
\times Y_{lm_l}^{*}(\Omega_{a}) Y_{l'm_{l'}}^{*}(\Omega_{b})  
<lm_l\frac{1}{2}m_{s_{a}} \mid jm> < j m j' m' \mid J_1 M_1 >
\nonumber \\
& & \times   <l'm_{l'}\frac{1}{2}m_{s_{b}}\mid j'm'>
<J_R M_R J_1 M_1 \mid J M > 
\nonumber \\
& & \times 
\mid (hh')^{-1} E_x J_R ; (p(\epsilon _alj)p'(\epsilon _bl'j')) J_1 ; JM> \;.
\label{wave}
\end{eqnarray}
The model that we adopt to account for the FSI is thus based on a
partial wave expansion in terms of two-hole-two-particle 
(2h2p) eigenstates of a mean-field
potential.  These wavefunctions are defined according to
\begin{eqnarray}
& & \mid (hh')^{-1} E_x J_R  ; (p(\epsilon _alj)p'(\epsilon _bl'j')) J_1 ;
JM> =
\sum_{mm'}\sum_{M_1M_R} 
\sum_{m_hm_{h'}} \frac{1}{\sqrt{1+\delta_{hh'}}}
\nonumber \\ & &
\times <jmj'm' \mid J_1 M_1> 
<J_R M_R J_1 M_1 \mid J M > 
<j_hm_hj_{h'}m_{h'} \mid J_R M_R > 
\nonumber \\ & &
\times (-1)^{j_h+m_h+j_{h'}+m_{h'}}
 c_{ljm}^{\dagger}c_{l'j'm'}^{\dagger} 
c_{h-m_{h}}
c_{h'-m_{h'}} \mid \Psi_0 >\;.
\end{eqnarray}
The continuum or particle ($p$) eigenstates of the
mean-field potential are characterized by $p(\epsilon l j)$. 
The energy at which the partial waves $p(\epsilon l j)$ are calculated
is determined by the momentum of the emitted nucleon :
$\epsilon ^2=k^2/(2M_N)$.  The central and Coulomb phase shifts are
denoted by $\delta _l$ and $\sigma _l$.  Throughout this work we
consider isospin not to be a good quantum number, in the sense that at
the level of the single-particle states we discriminate between
protons and neutrons.  The single-particle wave functions are
constructed through a Hartree-Fock calculation with an effective
Skyrme type interaction \cite{Waro}.  
For the protons a Coulomb part is added to
the mean-field potential.

\begin{figure}
\centering
\epsfysize=8.cm
\epsffile{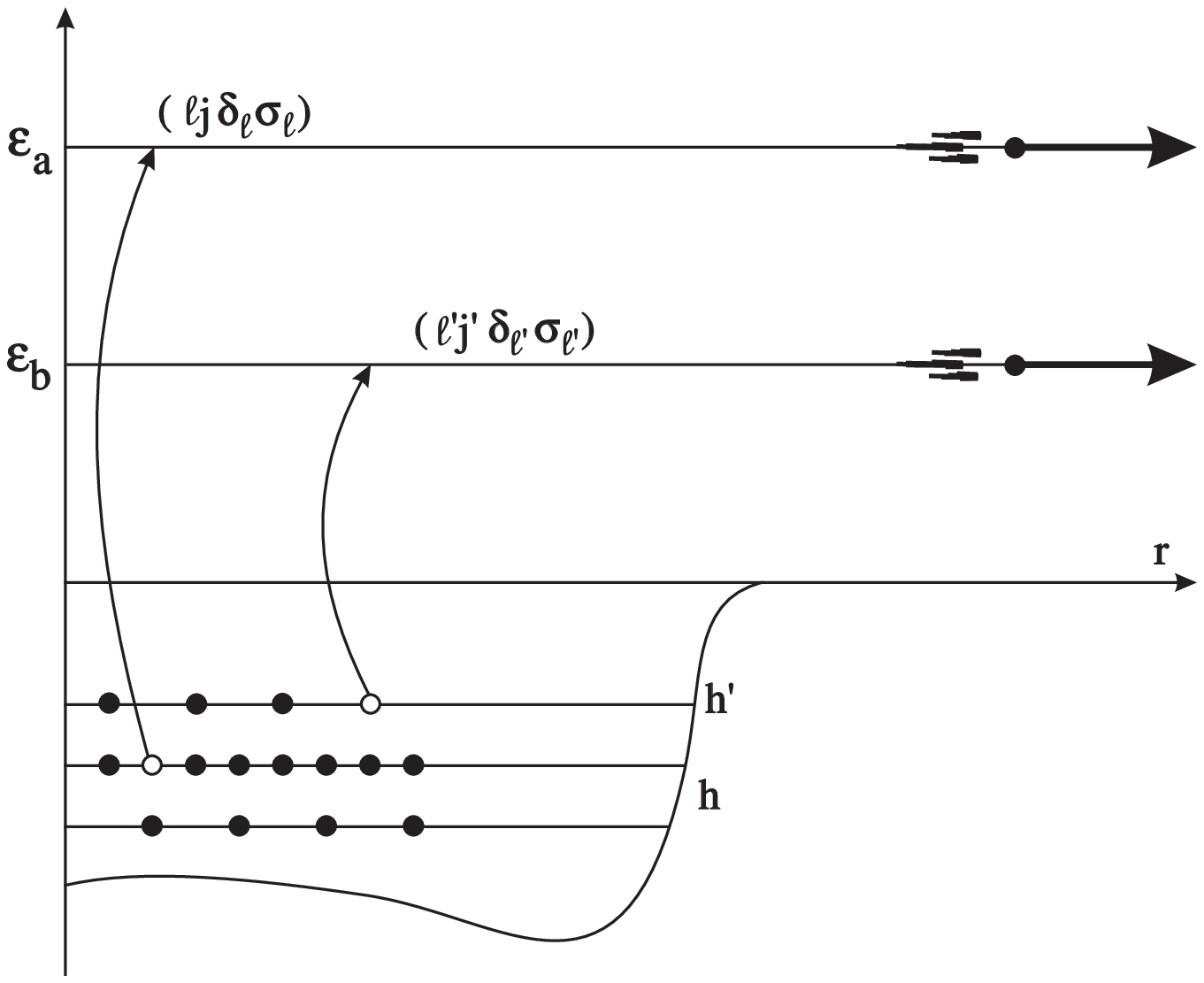}
\caption{Two-nucleon knockout as interpreted in a shell-model picture.
After ejection of two nucleons with energy $\epsilon _a$ and $\epsilon
_b$ the residual nucleus is left in a two-hole state (hh$'$)$^{-1}$.}
\label{fig:bosen1}
\end{figure}

The state in which the final state is created is determined
by $\mid (hh')^{-1} E_x J_R>$, with $E_x$ the excitation energy with
respect to the ground-state energy of the residual nucleus.  A schematic
drawing that illustrates the basic idea behind 
the expansion (\ref{wave}) is shown in Fig.~\ref{fig:bosen1}. It
should be noted that in reality only a fraction of the final-state
wave function will be of two-hole nature.  Indeed, the many-body
character of the residual system will make the two-hole strength to be
fragmented over a wide energy range.  In order to account for
this nuclear structure effect, 
``spectroscopic
factors'' have to be considered when calculating cross sections for
decay to specific states.  
The nuclear-structure aspects which are expected to play a
role in the $^{16}$O(e,e$'$pp) reaction have been the subject of a
recent study reported in Ref.~\cite{geurts}.       
 
At a first sight, the usefulness of Eq.~(\ref{wave})
is questionable as all expansions extend to infinity.  When performing
the calculations in coordinate space, finite
expressions for the transition matrix elements can be obtained by
performing an angular decomposition of the transition operators
$(J_{\lambda}(q),\rho(q))$.  This leads to the well-known expansion of
the nuclear four-current operator in terms of the Coulomb ($M_{JM}^{coul}$),
electric ($T_{JM}^{el}$) and magnetic ($T_{JM}^{mag}$) multipole
operators ($\widehat{J} \equiv \sqrt{2J+1}$)
\begin{eqnarray}
J_{\pm 1}(q)
&=&
- \sqrt{2\pi} \sum_{J \geq 1} i^J \widehat{J}  
\left(T_{J\pm 1}^{el}(q) \pm  T_{J\pm 1}^{mag}(q) \right)
\label{opel}
\\ 
\rho(q)
&=&
\sqrt{4\pi} \sum_{J \geq 0} i^J \widehat{J}  
M_{J0}^{coul}(q) \; .
\label{opcoul}
\end{eqnarray}
After inserting the expansions for the final state (\ref{wave}) and the
nuclear charge-current operator (\ref{opel}),
the transition matrix elements of Eqs. (\ref{mfia}) and (\ref{mfib}) can
be cast in the closed form :
\begin{eqnarray}
& & m_F^{fi} (\lambda = \pm 1)
 =  -\sqrt{2 \pi} \sum _{J \geq 1} i^J \hat{J} 
\sum_{lm_ljm}\sum_{l'm_{l'}j'm'} \sum_{J_1M_1} 
\nonumber \\ 
& & \times (4 \pi)^2 (-i)^{l+l'} 
\frac {\pi} {2 M_N \sqrt{k_a k_b}} e^{-i(\delta_l+\sigma_l
+\delta_{l'}+\sigma_{l'})}
  \nonumber \\
& & \times Y_{lm_l}(\Omega_{a}) Y_{l'm_{l'}}(\Omega_{b}) 
<lm_l\frac{1}{2}m_{s_{a}} \mid
jm>  <l'm_{l'}\frac{1}{2} m_{s_{b}} \mid j'm'> \nonumber \\
& & \times < j m j' m' \mid J_1 M_1 >
\frac{(-1)^{J_R-M_R+1}}{\hat{J_1}} 
<J_R -M_R J \lambda \mid J_1 M_1> \nonumber \\
& & \times [{\cal M}_{pp';hh'}^{el}(J_1,J,J_R)
+ \lambda  {\cal M}_{pp';hh'}^{mag}(J_1,J,J_R)     ] 
\label{feynb}
\end{eqnarray}
\begin{eqnarray}
& & m_F^{fi} (\lambda = 0)
 =  
\sqrt{4 \pi} \sum _{J \geq 0} i^J \hat{J} 
\sum_{lm_ljm}\sum_{l'm_{l'}j'm'} \sum_{J_1M_1} 
\nonumber \\
& & \times (4 \pi)^2 (-i)^{l+l'} 
\frac {\pi} {2 M_N \sqrt{k_a k_b}} e^{-i(\delta_l+\sigma_l
+\delta_{l'}+\sigma_{l'})}
  \nonumber \\
& & \times Y_{lm_l}(\Omega_{a}) Y_{l'm_{l'}}(\Omega_{b}) 
<lm_l\frac{1}{2}m_{s_{a}} \mid
jm>  <l'm_{l'}\frac{1}{2}m_{s_{b}}\mid j'm'> \nonumber \\
& & \times < j m j' m' \mid J_1 M_1 >
\frac{(-1)^{J_R-M_R+1}}{\hat{J_1}} 
<J_R -M_R J 0 \mid J_1 M_1> \nonumber \\
& & \times {\cal M}_{pp';hh'}^{coul}(J_1,J,J_R)\;,
\label{feynd}
\end{eqnarray}
where the matrix elements ${\cal M}$ have been defined according to :
\begin{eqnarray}
& & {\cal M}_{pp';hh'}^{el,mag}(J_1,J,J_R)  = 
<p(\epsilon_blj) p'(\epsilon_al'j');J_1
\| T^{el,mag}_{J} (q) \|hh' ; J_R > \nonumber \\
&& - (-1)^{j_h+j_{h'}+J_R} <p(\epsilon_b l j) p' (\epsilon_a l' j');J_1
\| T^{el,mag}_{J} (q) \|h'h ; J_R >\;, 
\label{mfel}
\\
& & {\cal M}_{pp';hh'}^{coul}(J_1,J,J_R)  = 
<p(\epsilon_blj) p'(\epsilon_al'j');J_1
\| M^{coul}_{J} (q) \|hh' ; J_R > \nonumber \\
&& - (-1)^{j_h+j_{h'}+J_R} <p(\epsilon_b l j) p' (\epsilon_a l' j');J_1
\| M^{coul}_{J} (q) \|h'h ; J_R >\;.
\label{mfcoul}
\end{eqnarray}
It should be stressed that the above expressions are general and that
no assumption whatsoever regarding the nature of the inital photoabsorption
mechanisms has been adopted.  The contributing photoabsorption
mechanisms will be discussed  in Sects. \ref{sec:mec} and \ref{sec:src}. 

\subsection{The semi-exclusive (e,e$'$N) cross section}
\label{sec:semi}
In comparison with exclusive (e,e$'$NN) reactions the semi-exclusive
(e,e$'$p) channel is more attractive from the experimental point of
view.  For model calculations, however, the semi-exclusive channel can
be more challenging than exclusive 2N knockout calculations.  Indeed
two-nucleon knockout is expected to feed the semi-exclusive (e,e$'$p)
channel as soon as the thresholds are crossed. The calculation of the
two-nucleon knockout contribution to the semi-exclusive channel
involves integration over the phase space of the undetected escaping
nucleon (either a proton or a neutron).  
In what follows we will exploit the fact that most of the
derivations for the two-nucleon knockout cross sections 
in previous sections are done in coordinate space to perform
some of these integrations analytically.  In this way we facilitate the
calculation of semi-exclusive cross sections enormously and develop a
framework that can be further exploited to investigate the role of the
dynamical nucleon-nucleon correlations in the semi-exclusive channel. As
most of the dynamical correlations are believed to be of two-body
nature, one could expect that they manifest themselves most clearly in
the two-nucleon emission channel.  It is a major challenge to trace
those kinematical regions where the semi-exclusive channel is
predominantly fed by emission of two nucleons.  Such investigations
start from considering the contribution from (e,e$'$pp) and (e,e$'$pn)
to the semi-exclusive channel~:
\begin{eqnarray}
& & {d^4 \sigma \over dE_p d \Omega _p d\epsilon ' d \Omega _{\epsilon '}}
(e,e'p)  =   \int d \Omega_{p'}
\int dE_{p'} {d^6 \sigma \over dE_p d \Omega _p dE_{p'} d \Omega _{p'}
d\epsilon ' d \Omega _{\epsilon '}}
(e,e'pp) \nonumber \\ & & +
\int d \Omega_{n}
\int dE_{n} {d^6 \sigma \over dE_p d \Omega _p dE_{n} d \Omega _{n}
d\epsilon ' d \Omega _{\epsilon '}}
(e,e'pn)\;.
\label{coneep}
\end{eqnarray}

An obvious way of determining the right-hand side of the above
expression is calculating the 2N knockout cross sections for a grid of
kinematical conditions that covers the full phase spaces $d \Omega
_{p'} d E _{p'}$ ((e,e$'$pp) contribution) and $d \Omega _{n} d E
_{n}$ ((e,e$'$pn) contribution).  Those cross sections can then be
integrated to obtain the semi-exclusive strength that is attributed to
2N knockout.  Such a procedure would be cumbersome and would consume a
lot a computing time. We present an
alternative method that avoids those numerical integrations at a very
small cost for the accuracy of the calculated cross sections.  We
start with remarking that most of the two-nucleon emission strength
$d^6 \sigma \over d E_p d\Omega _p dE_{a} d \Omega _{a} d\epsilon ' d
\Omega _{\epsilon '}$  (a=p$'$,n) is confined to a relatively small
fraction of the complete $d \Omega _a d \Omega _p$ phase space~: ever
since the original work by K.~Gottfried \cite{Got58} it has been clear
that 2N knockout reactions predominantly occur in back-to-back
situations (quasi-deuteron kinematics). 
Consequently, for a particular $dE _p d \Omega _p$
(semi-exclusive situation) most of the (e,e$'$ N$_a$ p) strength
will reside in
a restricted angular range for the second nucleon N$_a$ that remains
undetected.  In this restricted $d \Omega _a$ area the value of the
undetected nucleon momentum, which is determined from energy-momentum
conservation, will vary very slowly so that to a good approximation it
can be replaced by an ``average'' value $k _ a ^{ave}$. This average
momentum can be determined by imposing quasi-deuteron kinematics
(which is equivalent with considering the situation with 
zero recoil momentum ) :
\begin{equation}
\vec q - \vec k_a^{ave} - \vec k_p = \vec 0 \; ,
\end{equation}
where $\vec k_p$ is the momentum of the detected proton.
After introducing this ``average momentum'',
the integration over $d \Omega _{p'}$ and $d \Omega _n$
in Eq.~(\ref{coneep}) can be performed analytically. Indeed, after
inserting the expressions (\ref{feynb}) and (\ref{feynd}) into the
structure functions $W(\theta _a, \phi _a, \theta _b, \phi _b)$ of
Eqs. (\ref{wl}-\ref{wtt}) it can be shown that :
\begin{eqnarray}
& & \int d \Omega _a W_{L} (\theta _a, \phi _a, \theta _b , \phi _b)   = 
\sum _{J,J' \geq 0} \sum _ {lj} \sum _{l'j'} \sum _{l_1' j_1'} \sum
_{J_1 J_1'} \sum _{J_2} \sum _{hh' J_R} 
\frac {1} {1+ \delta _{hh'}} \frac {64 \pi ^6} {M_N^2 k_a^{ave} k_b}
\nonumber \\ 
& & \times {\cal B}\left(p(k_a
^{ave} lj),p'(k_b l'j'),p_1' (k_b l_1' j_1'), h, h', J_1, J, J_1
',J',J_R, J_2 \right) \nonumber \\ & & \times (-1)^{j+\frac {1} {2} +
J_R + j' -j_1'+J_2}
\left< J' \; 0 \; J \; 0 \mid J_2 \; 0 \right>
P_{J_{2}} \left( cos \theta _b \right)
\nonumber \\
& & \times {\cal M}_{pp';hh'}^{coul}(J_1,J,J_R)
\left({\cal M}_{pp_1';hh'}^{coul}(J_1',J',J_R)\right)^*
\label{eq:first}
\\
& & \int d \Omega _a W_{T}  (\theta _a, \phi _a, \theta _b , \phi _b) = 
\sum _{J,J' \geq 1} \sum _ {lj} \sum _{l'j'} \sum _{l_1' j_1'} \sum
_{J_1 J_1'} \sum _{J_2} \sum _{hh' J_R} 
\frac {1} {1+ \delta _{hh'}} \frac {32 \pi ^6} {M_N^2 k_a^{ave} k_b}
\nonumber \\
& & \times {\cal B}
\left[p(k_a ^{ave} lj),p'(k_b l'j'),p_1' (k_b l_1' j_1'), h, h',
J_1, J, J_1 ',J',J_R, J_2 \right] \nonumber \\
& & \times (-1)^{j-\frac {1} {2} + J_R + j' -j_1' + J + J'}
\left< J' \; -1 \; J \; 1 \mid J_2 \; 0 \right>
P_{J_{2}} \left( cos \theta _b \right)
\nonumber \\
& & \times \left\{ \left[ {\cal M}_{pp';hh'}^{el}(J_1,J,J_R)
\left({\cal M}_{pp_1';hh'}^{el}(J_1',J',J_R)\right) ^* \right. \right.
\nonumber \\
& & \left. +{\cal M}_{pp';hh'}^{mag}(J_1,J,J_R)
\left({\cal M}_{pp_1';hh'}^{mag}(J_1',J',J_R)\right)^* \right]
\times \left( 1 + (-1) ^ {J'+J+J_2} \right) \nonumber \\ 
& & + \left[ {\cal M}_{pp';hh'}^{el}(J_1,J,J_R)
\left({\cal M}_{pp_1';hh'}^{mag}(J_1',J',J_R)\right)^* \right.
\nonumber \\
& & \left. \left. +{\cal M}_{pp';hh'}^{mag}(J_1,J,J_R)
\left({\cal M}_{pp_1';hh'}^{el}(J_1',J',J_R)\right)^* \right]
\left( 1 + (-1) ^ {J'+J+J_2+1} \right) \right\} \\ 
\label{eq:second}
& & \int d \Omega _a W_{LT}  (\theta _a, \phi _a, \theta _b , \phi _b)  = 
\sum _{J \geq 1, J' \geq 0} \sum _ {lj} \sum _{l'j'} \sum _{l_1' j_1'} \sum
_{J_1 J_1'} \sum _{J_2 \geq 1}  \sum _{hh' J_R} 
\frac {1} {1+ \delta _{hh'}} \frac {64  \sqrt{2}\pi ^6} {M_N^2 k_a^{ave} k_b}
\nonumber \\
& & \times Re \Biggl[ {\cal B}\left[p(k_a ^{ave} lj),p'(k_b l'j'),p_1' (k_b l_1'
j_1'), h, h', J_1, J, J_1 ',J',J_R, J_2 \right] \nonumber \\ & & \times
(-1)^{j_1+\frac {1} {2} + J_R + j' -j_1'+J_2}
\left< J \; 1 \; J' \; 0 \mid J_2 \; 1 \right>
\frac {1} {\sqrt{J_2 (J_2+1)}}
P_{J_{2}}^1 \left(cos \theta _b \right)
\nonumber \\
& &  \times \left\{ {\cal M}_{pp';hh'}^{el}(J_1,J,J_R)
\left({\cal M}_{pp_1';hh'}^{coul}(J_1',J',J_R)\right)^* 
\left( e^{i \phi _b} + (-1) ^ {J + J' + J_2} e ^ {- i \phi _ b} \right)
\right. \nonumber \\
& & +  {\cal M}_{pp';hh'}^{mag}(J_1,J,J_R)
\left({\cal M}_{pp_1';hh'}^{coul}(J_1',J',J_R)\right)^*
\nonumber \\
& & \left. \times
\left( e^{i \phi _b} + (-1) ^ {J + J' + J_2+1} e ^ {- i \phi _ b}
\right) \right\} \Biggr]
 \\
\label{eq:thirth}
& & \int d \Omega _a W_{TT}  (\theta _a, \phi _a, \theta _b , \phi _b)  = 
\sum _{J,J' \geq 1} \sum _ {lj} \sum _{l'j'} \sum _{l_1' j_1'} \sum
_{J_1 J_1'} \sum _{J_2 \geq 2} \sum _{hh' J_R} 
\frac {1} {1+ \delta _{hh'}} \frac {32  \pi ^6} {M_N^2 k_a^{ave} k_b}
\nonumber \\ 
& & \times {\cal B}
\left[p(k_a ^{ave}lj),p'(k_b l'j'),p_1' (k_b l_1' j_1'), h, h',
J_1, J, J_1 ',J',J_R, J_2 \right] \nonumber \\
& & \times (-1)^{j-\frac {1} {2} + J_R + j' -j_1'+J_2}
\left< J' \; 1 \; J \; 1 \mid J_2 \; 2 \right>
\frac {1} {\sqrt{(J_2-1) J_2 (J_2+1)
(J_2+2)}} P_{J_{2}}^2 \left(cos \theta _b \right)
\nonumber \\
& & \times \Biggl[ \biggl[ {\cal M}_{pp';hh'}^{el}(J_1,J,J_R)
\left({\cal M}_{pp_1';hh'}^{el}(J_1',J',J_R)\right)^* 
\nonumber \\
& & - {\cal M}_{pp';hh'}^{mag}(J_1,J,J_R)
\left({\cal M}_{pp_1';hh'}^{mag}(J_1',J',J_R)\right)^* \biggr]
\nonumber \\
& & \times \left( e^{- 2 i \phi _b} + (-1) ^ {\left( J_2 + J' + J \right)}
e^{ 2 i \phi _b} \right) 
\nonumber \\ & &
-2 Im \biggl[  {\cal M}_{pp';hh'}^{mag}(J_1,J,J_R)
\left({\cal M}_{pp_1';hh'}^{el}(J_1',J',J_R)\right)^* \biggr]
\nonumber \\
& & \times \left( e^{- 2 i \phi _b} - (-1) ^ {\left( J_2 + J' + J \right)}
e^{ 2 i \phi _b} \right) \Biggr]
\; ,
\label{eq:fourth}
\end{eqnarray}
where the function ${\cal B}$ was defined according to :
\begin{eqnarray}
{\cal B} & & \left[p(k_a lj),p'(k_b l'j'),p_1' (k_b l_1' j_1'), h, h',
J_1, J, J_1 ',J',J_R, J_2 \right]
= \nonumber \\
& &  i^{J-J'+l_1'-l'} e^{-i \left( \delta _{l'} + \sigma _{l'}
    -\delta _{l_{1}'} - \sigma _{l_{1}'} \right) } 
\widehat{J} \widehat{J'} \widehat{j'} \widehat{j_1 '} 
\widehat{J_1} \widehat{J_{1}'} 
 \frac {1} {2} \left[ 1 + (-1)^
{ \left( J_2 + l'+ l_1' \right) } \right]
\nonumber \\
&& \times \left\{ \begin{array}{lll}
    j' \; j_1' \; J_2  \nonumber \\
    J_1 ' \; J_1 \; j  
    \end{array} \right\} 
\left\{ \begin{array}{lll}
    J_1 \; J \; J_R   \\
    J' \; J_1 ' \; J_2  
    \end{array} \right\}
\left< j' \frac{1}{2} j_1 ' - \frac {1} {2} \mid J_2 0 \right>
\end{eqnarray} 
These expressions allow to compute directly the semi-exclusive cross
sections as a function of the kinematical variables ($\theta _p, \phi
_p, T_p$) without explicitly calculating the intermediate (e,e$'$NN) 
angular cross sections.  In the actual calculations we account for the
spreading of the two-hole strength in the energy spectrum of the A-2
system.  This is done with the aid of a model explained in
Ref.~\cite{ryc94}.

\subsection{Meson-exchange and isobaric currents}
\label{sec:mec}
In an independent-particle picture, the most direct source of
two-nucleon knockout strength would be the coupling of the
electromagnetic field to the hadronic two-body currents.  These
currents are mediated by the mesons which lie at the basis of the
one-boson exchange picture of the nucleon-nucleon interaction.  Here,
we do not make any attempt to consider medium-corrections for the
meson fields.  In constructing the two-body currents, our starting
point is the one-boson exchange (OBE) picture of the nucleon-nucleon
interaction as it has been derived from nucleon-nucleon scattering
data.   The long-range part of the OBE potential is well
established and dominated by pion exchange.  In coordinate space it takes on
the well-known form as a sum of a spin-spin and tensor interaction
term \cite{mach}
\begin{eqnarray}
V _ {\pi} (\vec r _{12} ) & = & \vec{\tau} _1 . \vec{\tau} _2 
\frac {1} {3}
\left( \frac {f_{\pi NN}}  {m_ {\pi}} \right) ^2
\nonumber \\
& & \times \left\{ \vec{\sigma} _1 .  \vec{\sigma} _2 + S_{12}(\hat{r})
\left[ 1+ 
          \frac {3} {m_ {\pi} r } + 
	  \frac {3}  {(m_ {\pi} r)^2} 
\right] \right\}
\frac {e^ {-m_{\pi}r}}  {r} \; .
\end{eqnarray} 
The current operator satisfying the continuity equation with this
potential can be constructed from general field-theoretical methods.
We have considered pseudovector $\pi NN$ coupling and treated the
two-body currents in the non-relativistic limit \cite{marc}.   This
results in the well-known $\pi$-exchange current operators :
\begin{eqnarray}
\label{eq:pisea}
\vec J_{\pi} (\vec q _1 , \vec q _2)  & = & (-i)e 
{{f_{\pi NN}^2} \over {m_\pi ^2}}
(\vec \tau _1 \times \vec \tau _2)_3
  \left( 
{{\vec \sigma _1 \left( {\vec
   \sigma _2 \cdot \vec q_2} \right) } \over {\vec q_2 ^2 + m_\pi ^2}}
 \; - \;
{{\vec \sigma _2 \left( {\vec 
   \sigma _1 \cdot \vec q_1} \right)} \over  {\vec q_1 ^2 + m_\pi ^2}}
\right. \nonumber \\
& & \left. - { {\left( \vec \sigma _1 \cdot \vec q _1 \right) 
     \left( \vec \sigma _2 \cdot \vec q _2 \right)
    } \over 
    {\left( \vec q_1^2 + m_\pi ^2 \right)
     \left( \vec q_2^2 + m_\pi ^2 \right)
    } }
    \;(\vec q _1\,-\,\vec q_2) \right) \;,
\end{eqnarray}
where the first two terms refer to the so-called Seagull current and
the last term, which is quadratic in the pion propagators, to the
pion-in-flight current.

\begin{figure}
\centering
\epsfysize=6.cm
\epsffile{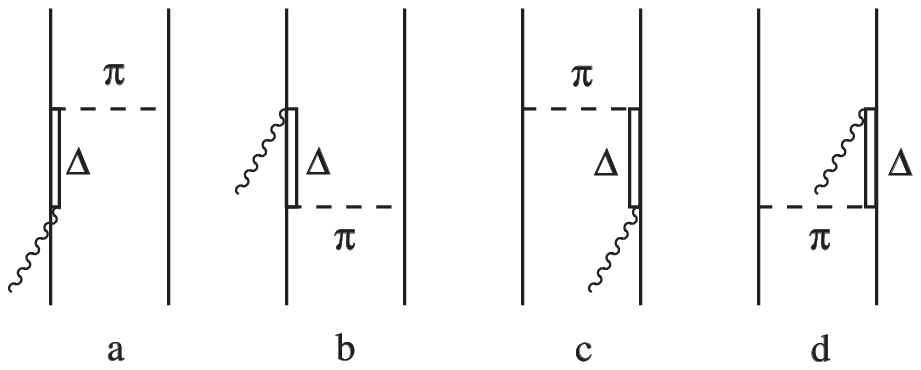}
\caption{The four diagrams corresponding with $\Delta _{33}$ creation
and $\pi$-exchange.}
\label{fig:delta}
\end{figure}

In addition to the meson-exchange currents (MEC) derived from the OBE
potential we have also considered processes in which 
nucleon excitations $\Delta _{33}$ are created (Fig.~\ref{fig:delta}).
Here, the model
dependency is intrinsically larger as the (transverse) currents cannot
be constrained by the continuity equation.  In order to construct the
currents associated with $\Delta _{33}$ creation with subsequent pion
exchange, we also rely on field-theoretical methods and adopt a
non-relativistic approach.  The $\pi N \Delta$ coupling
is considered in the standard form~: 
\begin{equation}
{\cal L} _{\pi N \Delta} = \frac {f_{\pi N \Delta}} {m _{\pi}}
\left( \vec{S} ^{\dagger} . \vec{\nabla} \right)
\left( \vec{T} ^{\dagger} . \vec{\pi} \right) \; ,
\end{equation}
where $\vec{S}$ and $\vec{T}$ denote the spin and isospin $\frac {1}
{2} \rightarrow \frac {3} {2}$ transition operators.
In the $\gamma N \Delta$ coupling we retain solely the magnetic term
\begin{equation}
{\cal L} _{\gamma N \Delta} = F_{\Delta}(q_\mu^2)
\frac {f_{\gamma N \Delta}} {m _{\pi}}
\left( \vec{S} ^{\dagger} \times \vec{\nabla} \right) . \vec{A}
\vec{T} _{3} ^{\dagger} \; ,
\end{equation}
with $\vec{A}$ the external electromagnetic field and
$F_{\Delta}(q_\mu^2)$ the electromagnetic form factor of the delta for
which we used the standard dipole form.  The (marginal)
electric quadrupole term in the $\gamma N \Delta$ coupling 
is neglected throughout this work.  With the above coupling lagrangians
we arrive at the following expression for the $\Delta _{33}$-current
with $\pi$-exchange :
\begin{eqnarray}
\label{eq:pidelta}
&&\vec J_{\pi \Delta} (\vec q,\vec q _1, \vec q _2) =
{i \over 9}{{f_{\gamma N\Delta }f_{\pi NN}f_{\pi
N\Delta }} \over {m_\pi ^3\,}} F_{\Delta}(q_\mu^2) \;\nonumber\\ 
&&\times\; \Biggl\{ {\left[
G_{\Delta}^{res} + G_{\Delta}^{non-res} 
\right]}
\,\nonumber\\
&&\quad\times\;\left[ {\,4\,\left( {\vec \tau _2} \right)_3\;\left(
{\vec q_2\times
\vec q} \right)\,{{\vec \sigma _2\cdot \vec q_2} \over {\vec{q_2}^2+m_\pi
^2}}+4\,\left( {\vec \tau _1} \right)_3\;\left( {\vec q_1\times \vec q}
\right)\,{{\vec \sigma _1\cdot \vec q_1} \over {\vec{q_1}^2+m_\pi ^2}}}
\right.\nonumber\\
&&\left. {\left.\quad {+\,(\vec \tau _1\times \vec \tau _2)_3\,\left[
{\left( {\vec \sigma _2\times \vec q_1} \right){{\vec \sigma _1\cdot
\vec q_1} \over {\vec{q_1}^2+m_\pi ^2}}-\left( {\vec \sigma _1\times \vec
q_2} \right){{\vec \sigma _2\cdot \vec q_2} \over {\vec{q_2}^2+m_\pi
^2}}} \right.} \right]\times \vec q} \right]\nonumber
\\ && +\,\left[
G_{\Delta}^{res} - G_{\Delta}^{non-res} 
\right]\nonumber\\
&&\quad\times\;\left[ {-2\,i\,\,\left( {\vec \tau _2}
\right)_3\;\left( {\left( {\vec \sigma _1\times \vec q_2}
\right)\times \vec q} \right){{\vec
\sigma _2\cdot \vec q_2} \over {\vec{q_2}^2+m_\pi ^2}}}
\right.
\nonumber \\
& & -2\,i\,\,\left( {\vec \tau _1} \right)_3\;\left( {\left( {\vec
\sigma _2\times \vec q_1} \right)\times \vec q} \right){{\vec \sigma
_1\cdot \vec q_1} \over {\vec{q_1}^2+m_\pi ^2}}\nonumber\\
&&\quad  {\left. {\left. {-\,2\,i\,\,(\vec \tau _1\times \vec \tau
_2)_3\,\left[ {\vec q_2{{\vec \sigma _2\cdot \vec q_2} \over {\vec{
q_2}^2+m_\pi ^2}}-\vec q_1{{\vec \sigma _1\cdot \vec q_1} \over {\vec{
q_1}^2+m_\pi ^2}}} \right.} \right]\times \vec q} \right]} \Biggr\} \;.
\end{eqnarray}
The $\pi N \Delta$ and $\gamma N \Delta$ coupling constants are taken
from ref.\cite{Oset} : ${f^2}_{\pi N \Delta}$ / 4 $\pi$ = 0.37 , 
$f_{\gamma N\Delta}$ = 0.12.  In the above expression for the isobaric
current $G_{\Delta}^{res}$ denotes the $\Delta _{33}$ propagator for
the resonant diagrams (Fig.\ref{fig:delta}(a) and (c)).  The
corresponding propagator for the so-called non-resonant diagrams  
(Fig.\ref{fig:delta}(b) and (d)) is denoted by $G_{\Delta}^{non-res}$.   
As we will consider energy transfers in
the resonance region, special attention has been paid to constructing
the $\Delta _{33}$ propagators.  A free $\Delta _{33}$ excitation
obtains a width through $\pi N$ decay.  The corresponding propagator
would then read 
\begin{equation}
G_{\Delta}^{res} = {{1} 
                    \over
                    {-E_{\Delta}^{res}+M_{\Delta} - {i \over 2} \Gamma
                     _{\Delta}^{res}}} \; ,
\label{eq:freedelta}
\end{equation}
where $E_{\Delta}^{res}$ is the intrinsically available energy for the
resonance and the width  $\Gamma _{\Delta}^{res}$ becomes
\cite{Oset,dekker} 
\begin{equation}
\Gamma _{\Delta}^{res} = \frac {2} {3} \frac {f _{\pi N \Delta} ^2} {4
\pi} \frac { \mid \vec{p} _{\pi} \mid ^3 } {m _{\pi} ^2}
\frac {M_N} {\sqrt{s}} \; ,
\end{equation}
where $\vec{p} _{\pi}$ is the decay momentum in the center-of-mass (c.o.m)
frame of the $\pi N$ system and $\sqrt{s}$ the total c.o.m. energy
of the pion and nucleon.  In the medium, however, the $\pi N$ decay
will be blocked by the Pauli principle.  Various pion-nucleus
\cite{osterfeld}, photoabsorption \cite{bianchi} and inclusive (e,e$'$)
\cite{koch,oconnell} experiments have pointed towards other strong
medium modifications of the $\Delta _{33}$ resonance.  Indeed, the
inclusive A(e,e$'$) spectra show a pronounced broadening and damping
of the resonance in comparison with A times the free electron nucleon
cross section.  It is common belief that this is mainly due to the
coupling of the dominant $\Delta \rightarrow \pi N$ decay with the
$\Delta N \rightarrow NN$ channel.  It is precisely the latter channel
which is under investigation in two-nucleon knockout experiments.  The
$\Delta$-hole model \cite{koch} 
has been applied successfully to the description
of inclusive (e,e$'$) reactions in the $\Delta _{33}$ resonance region
\cite{oconnell}. In the $\Delta$-hole model one adopts a dynamical
description of isobar propagation in the nucleus.  This results in a
propagator of the form \cite{koch}
\begin{equation}
G_{\Delta}^{res} = {{1} 
                    \over
                    {-E_{\Delta}^{res}+M_{\Delta} - {i \over 2} \Gamma
                     _{\Delta}^{res} + \delta W + W_{\pi} + W_{sp}}} \; ,
\end{equation}
where $\delta W$ accounts for Pauli blocking and $W_{\pi}$ for
coupling of the resonance to $\pi^o$ and the nuclear ground state.
These two terms tend to cancel each other so that $W_{sp}$ will induce
the major correction with respect to the free delta propagator of
Eq.~(\ref{eq:freedelta}).  The
$W_{sp}$ is a semi-phenomenological parametrization of the coupling of
the delta with the more complicated channels.  In the orginal
formulation of the delta-hole model the $W_{sp}$ was chosen to have a
density-dependent 
central and spin-orbit part.  Chen and Lee \cite{chen} have shown that
an equally good description of the $^{12}$C(e,e$'$) cross sections
could be obtained by simply assuming that
\begin{equation}
W_{sp} [\mathrm{MeV}] = -30-40 \; \mathrm{i} \; . 
\label{eq:chen}
\end{equation} 
This prescription is in agreement with the results from recent total
photoabsorption measurements \cite{bianchi} from which it was
concluded that the $\Delta$ mass and width increases with growing
nuclear density.  The values for the $\Delta$  mass and width for
$^{12}$C quoted in
Ref.~\cite{bianchi} are in agreement with the above prescription.     
In what follows we will adopt the procedure (\ref{eq:chen}) 
to account for the
medium effects on the $\Delta _{33}$ propagation.  Recently,
this approach has
been applied with some success to the description of the 
photonenergy dependence of the
$^{12}$C($\gamma$,pp) and ($\gamma$,pn) cross sections in the
resonance region \cite{douglas}. 

It is well known that the delta peak in the inclusive (e,e$'$) cross
section appears at a higher energy transfer $\omega$ than 300~MeV 
 which is typical for real photon induced
reactions.   It is convenient to introduce the equivalent
photon energy $K$ to produce the same nuclear excitation energy as the
virtual photon ($\vec{q},\omega$)
\begin{equation}
K=\omega + \frac {q_{\mu}q^{\mu}}{2M_N} \; .
\end{equation}   
A real photon with energy $K$ produces the same $\pi N$ c.o.m. energy
as a virtual photon $(\vec q,\omega)$.  
  
As we are dealing with off-shell nucleons the photon energy is not
completely available for internal excitation of the nucleon.  As
pointed out in Refs.\cite{thomas,reply} a reasonable substitution for
$E_{\Delta}^{res}$ is :
\begin{equation}
\left( E_{\Delta}^{res} \right) ^2 = 
      \left(M_N - \epsilon _h \right)^2 
+ 2 K (M_N - \epsilon _h) \; ,   
\end{equation}
where $\epsilon _h$ is the average binding energy of the nucleon on
which the pion is reabsorbed.  Remark that $\epsilon _h$ depends on
the shell in which the nucleon is residing.

For the non-resonant diagrams 
(Fig.\ref{fig:delta}(c) and (d)) the propagator becomes
\begin{equation}
G_{\Delta}^{non-res} = {{1} 
                    \over
                    {-E_{\Delta}^{non-res}+M_{\Delta} }} \; ,
\end{equation}
with
\begin{equation}
\left( E_{\Delta}^{non-res} \right)  = \sqrt{
\left( M_N + {{(T_a + T_b)} \over {2}} \right)^2 + \mid \vec q \mid ^2  } 
      - \omega   
\end{equation}

In the static limit (small $\omega$ and $q$), one obtains :
\begin{equation}
G_{\Delta}^{res} = G_{\Delta}^{non-res} = {{1} \over {M _{\Delta}
-M_N}} \; ,
\end{equation}    
in which case the isobaric current operator of 
Eq.~(\ref{eq:pidelta}) reduces to
the static operator form as e.g. derived in Ref.\cite{riska83}.  By no
means the static limit should be considered realistic as soon
as one is approaching equivalent photon energies $K$ that probe the
real  resonance region.  

In fitting nucleon-nucleon scattering data in terms of a particular
One-Boson exchange potential it is a common procedure to regularize
the $\pi NN$ vertices for the finite size of the hadrons and 
the complexity of physical mechanisms that
are thought to happen at 
distances which are short
as compared with the average range of the pion.
This is
frequently done by introducing a monopole hadronic form factor.
\begin{equation}
\frac
{\Lambda _{\pi}^2 - m_{\pi}^2} {\Lambda _{\pi}^2 + k^2} 
\end{equation}
at each $\pi NN$ vertex.  For the actual calculations, 
these hadronic form factors have also been
introduced in the meson-exchange and isobaric current operators of 
Eqs.~(\ref{eq:pisea}) and (\ref{eq:pidelta}).  Throughout this work, 
a pion cut-off mass $\Lambda _{\pi}$ of
1250 MeV has been used \cite{mach}.
 
\subsection{Ground-state correlations}
\label{sec:src}
In all of the above considerations we have adopted the independent
particle model (IPM) in which all nuclear wave functions are cast in a
Slater determinant form.  In this Subsection we present a method to
impose corrections to that picture.  To that purpose we rely on a
perturbation expansion method \cite{gaudin,clark} to calculate
transition matrix elements between many-particle wave functions that
have been corrected for correlation effects that go beyond the IPM.
Here, correlations in the nuclear wave functions are implemented via a
technique inspired by  
the so-called correlated basis function (CBF) theory \cite{fantoni} 
in which
correlated wave functions $\overline{\Psi}$ are derived from their
(uncorrelated) independent particle limit through the operation 
of an operator
$\widehat{{\cal G}}$. The latter corrects the Slater determinant $\Psi$ for
short-range and other correlations not accounted for in the IPM :
\begin{equation}
\mid \overline{\Psi} \rangle = \frac {\widehat{{\cal G}} \mid \Psi \rangle 
} {\sqrt{
\left< \Psi \mid \widehat{{\cal G}}^{\dagger}  \widehat{{\cal
G}}  \mid \Psi \right>}} 
\; .
\end{equation}
In the CBF theory, the operator $\widehat{{\cal G}}$ takes on the form
of a symmetrized product of two-body correlation operators :
\begin{equation}
\widehat{{\cal G}}= {\widehat{\cal S}} \Biggl[
 \prod _{i<j=1} ^{A} \sum _{p=1,6} 
f_{ij}^p(\vec{r}_{ij},\vec{R}_{ij}) 
\widehat{O}_{ij}^p \Biggr] \;,
\end{equation}
where ${\vec r}_{ij}={\vec r}_i - {\vec r}_j$, ${\vec R}_{ij}=({\vec r}_i +
{\vec r}_j)/2$ and ${\widehat{\cal S}}$ is the symmetrization operator.  The
$\widehat{O}_{ij}^p$ are the isoscalar and isovector 
isospin operators of the scalar, spin
and tensor type \cite{fabro} : 
\begin{eqnarray}
\widehat{O}_{ij}^p & \epsilon & \{ 1(p=1),\vec{\tau} _i  \cdot 
\vec{\tau} _j
(p=2), \vec{\sigma} _i \cdot \vec{\sigma} _j (p=3),  \\ & & 
\vec{\sigma} _i
\cdot \vec{\sigma} _j \nonumber  \vec{\tau} _i 
\cdot \vec{\tau} _j (p=4), S_{ij} (p=5), S_{ij} \vec{\tau} _i 
\cdot \vec{\tau} _j (p=6) \} \; .
\end{eqnarray} 
The different components in the operator $\widehat{{\cal G}}$ reflect
the fact that the nucleon-nucleon force is a function of the spin and
isospin orientation of the interacting particles. Of all of the above
components the central $(p=1)$ and tensor $(p=6)$ operator
have been
notified \cite{gua96,vijay} to induce the largest correlation 
corrections to the nuclear
Slater determinants obtained in the IPM approach.  

\begin{figure}
\centering
\epsfysize=9.cm
\epsffile{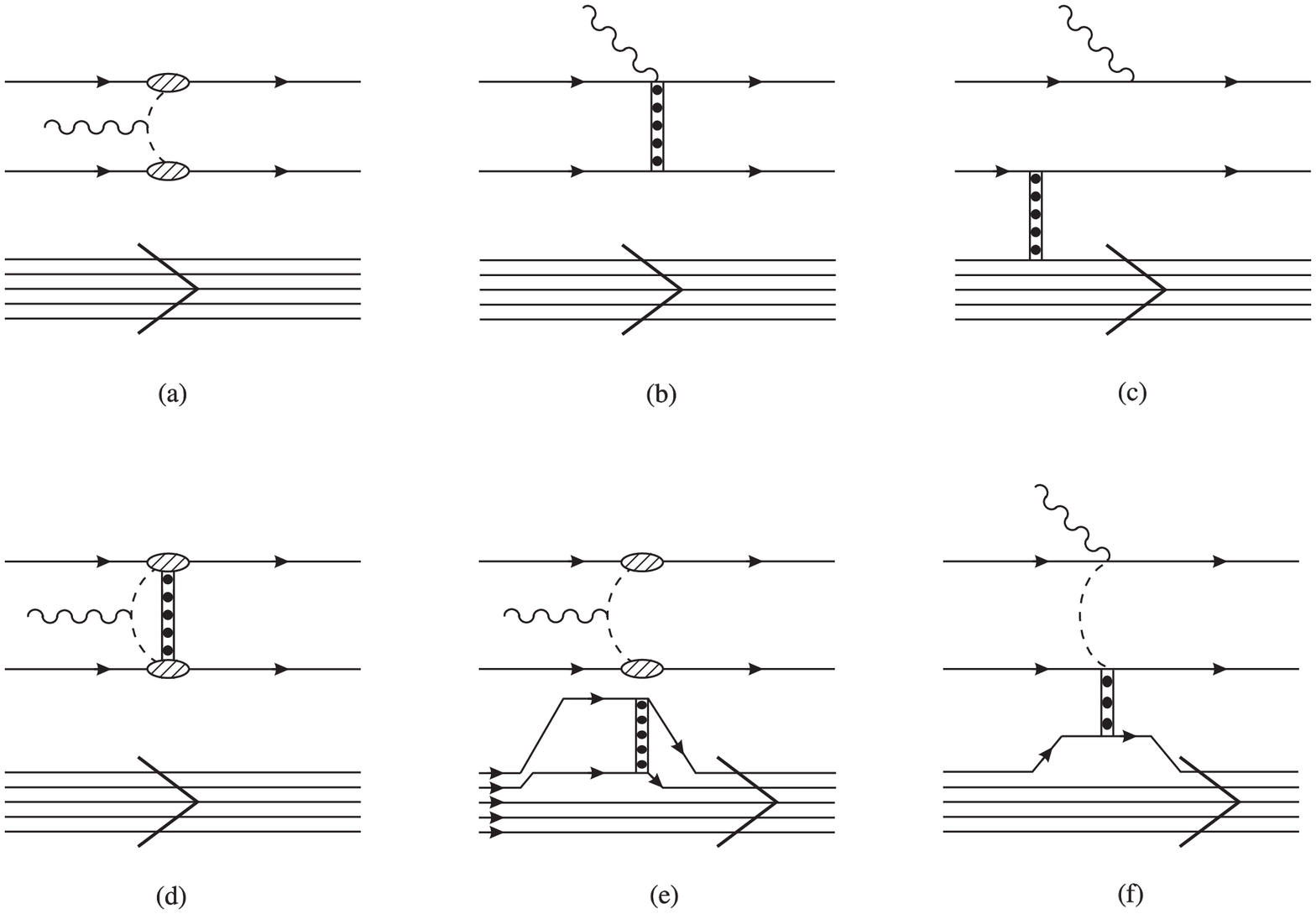}
\caption{
Different contributions to photoinduced two-nucleon knockout.  Meson
exchange processes are denoted with the dashed line.  The photon is
indicated with a wavy line and the ground-state correlations with the
heavy dots.}
\label{fig:bosen2}
\end{figure}

Throughout this paper we will restrict ourselves to the effects induced
by the central operator $(p=1)$, that induces short-range correlations
(SRC) to the
nuclear wave functions obtained in an IPM approach.  It should be
stressed, however, that the techniques expanded in this paper can also
be applied to the other terms.  In most of the
calculations that address the correlation function $f_{ij}^{p=1}({\vec
r}_{ij},{\vec R}_{ij})$ it is assumed that the functional dependence can be
restricted to the relative coordinate of the interacting nucleon pair
$r_{ij}=\mid {\vec r}_i - {\vec r}_j \mid$.  This assumption, which
considerably simplifies the calculation of the correlation functions,
can be justified by considering that the range of the correlation
function $f^{p=1}_{ij}$ is small with respect to the surface thickness
of the nucleus.  Accordingly, the dependence of $f^{p=1}_{ij}$ on the
c.o.m. coordinate ${\vec R}_{ij}$ is anticipated to be small. 
As a result of retaining solely the corrections induced by
the short-range correlations, the correlated nuclear wave functions used
throughout this paper read 
\begin{equation}
\overline{\Psi}({\vec x}_1,....,{\vec x}_A)=\prod _{i<j=1} ^{A}
f_{ij}^{p=1}(r_{ij}) \Psi({\vec x}_1,....,{\vec x}_A)
/ \sqrt{N} \; ,
\end{equation}
where $\vec x$ is a shorthand notation for the radial and spin coordinates
and $N$ is the normalization factor
\begin{eqnarray}
N & = & \int d \vec x_1  .... d \vec x_A \Psi ^\dagger 
({\vec x}_1,....,{\vec x}_A)
\left( \prod _{i<j=1} ^{A} f_{ij}^{p=1}(r_{ij}) \right)^*
\nonumber \\ 
& & \times \prod _{i<j=1} ^{A} f_{ij}^{p=1}(r_{ij}) 
\Psi ({\vec x}_1,....,{\vec x}_A) 
\end{eqnarray}
Within the adopted assumptions an arbitrary transition matrix element
between a correlated ground state $\mid \overline{\Psi} _o>$ 
and a final state $\mid \overline{\Psi} _f>$ can be rewritten as 
\begin{equation}
\left\langle \overline{\Psi} _f \mid \widehat{O} \mid \overline{\Psi} _o
\right\rangle \equiv \frac {1}  {\sqrt{N_i N_f}}
\left\langle \Psi _f \mid \widehat{O}^{eff} \mid \Psi _o 
\right\rangle \; ,
\label{cormat}
\end{equation}
with
\begin{equation}
\widehat{O}^{eff} = \prod _{i<j=1} ^{A} \left( 1-g(r_{ij})
\right)^{\dag} \widehat{O} \prod _{k<l=1} ^{A} \left( 1-g(r_{kl})\right)
\; ,
\label{effop}
\end{equation}
where we have introduced the central correlation function $g$ which is
defined according to $g(r_{ij})=1-f_{ij}^{p=1}(r_{ij})$.  In the absence
of short-range corrections the correlation function $g(r_{ij})$ would
simply be zero.   The effective operator approach as formally written in
Eq.~(\ref{cormat}) has the marked advantage that we can rely on standard
techniques, like second quantization, when calculating the transition
matrix elements between correlated many-body states.  
The adopted approach also allows to
put the short-range correlations on the same footing as other mechanisms
that can feed the electroinduced two-nucleon knockout channel.  In this
way it will become easier to evaluate the sensitivity of the calcaluted
cross sections to the SRC relative to the other mechanisms.  The
effective operator of Eq.~(\ref{effop}) is an A-body operator which
makes the exact calculation of ``correlated'' transition matrix elements
of the type (\ref{cormat}) not feasible.  Various cluster expansions,
however, have been developed to approximate the matrix elements of
Eq.~(\ref{cormat}) 
\cite{clark,hodgson}.  These techniques usually address matrix
elements of the type $< \overline{\Psi} _o \mid \widehat{O} \mid
\overline{\Psi}_o >$, where $\overline{\Psi} _o$ is the correlated
ground state of the nuclear system.  
Here, we are facing a quite different situation in the sense that
we are dealing with  transition matrix elements between wave functions 
of totally different origin, namely $\overline{\Psi} _o$ and a final
state with two nucleons residing in the continuum.  Given
that two-nucleon knockout calculations are already quite cumbersome in a
pure IPM, there is a practical limitation on the number of terms in
the cluster expansion that
can be accounted for.  This does not necessarily mean, however, that the
main effects of the SRC  cannot be calculated to a high degree of
accuracy.  Most of the cluster expansion techniques developed for the
calculation of the ``correlated'' matrix elements rely on an expansion
into the different orders of the central correlation function $g$. 
Here, we will restrict ourselves to the terms linear in the correlation
function $g$.  Higher order terms require three and more correlated
nucleons.  Given that the central correlation effect is very short
ranged, this type of multi-nucleon correlations could be expected to be
less important than two-body correlations at normal nuclear densities. 
In all forthcoming derivations we assume that the transition operator
$\widehat{O}$ in expression~(\ref{cormat}) has a one- and two-body part :
\begin{equation}
\widehat{O}=\sum_{i} \widehat{O}^{[1]}(i) + \sum_{i<j}
\widehat{O}^{[2]}(i,j) \; .
\end{equation}        
Within the adopted assumptions the effective transition operator can
then be written as :
\begin{eqnarray}
\widehat{O}^{eff} & =& \sum_{i} \widehat{O}^{[1]}(i) + \sum_{i<j}
\widehat{O}^{[2]}(i,j)
- \sum _{i<j} \biggl[ \widehat{O}^{[1]}(i)+\widehat{O}^{[1]}(j)
\biggr] g(r_{ij})
\nonumber \\
& & - \sum _{i<j<k} \biggl[ \widehat{O}^{[1]}(i)g(r_{jk})
+\widehat{O}^{[1]}(j)g(r_{ik})+\widehat{O}^{[1]}(k)g(r_{ij}) \biggr]
\nonumber \\
& & -\sum _{i<j} \widehat{O}^{[2]}(i,j)g(r_{ij})
\nonumber \\
& & - \sum _{i<j<k} \biggl[ \widehat{O}^{[2]}(i,j)g(r_{ik})
+ \widehat{O}^{[2]}(i,j)g(r_{jk})
+ \widehat{O}^{[2]}(i,k)g(r_{ij})
 \nonumber \\ 
& &  + \widehat{O}^{[2]}(i,k)g(r_{jk})
+ \widehat{O}^{[2]}(j,k)g(r_{ij})
+ \widehat{O}^{[2]}(j,k)g(r_{ik}) \biggr]
\nonumber \\
& & - \sum _{i<j<k<l} \biggl[ \widehat{O}^{[2]}(i,j)g(r_{kl})
+ \widehat{O}^{[2]}(i,k)g(r_{jl})
+ \widehat{O}^{[2]}(i,l)g(r_{jk})
 \nonumber \\
& &  + \widehat{O}^{[2]}(j,k)g(r_{il}) 
+ \widehat{O}^{[2]}(j,l)g(r_{ik})
+ \widehat{O}^{[2]}(k,l)g(r_{ij}) \biggr] \; .
\label{eq:heidi}
\end{eqnarray}
For the sake of brevity we did not write the terms in $g^{\dagger}$,
that refer to final-state correlations, in the
above expression.  Formally these contributions have exactly the same
form as the terms generated by the initial-state correlations.   
The above expression clearly illustrates how the short-range effects are
put on the same footing as the other contributions in the
photoabsorption mechanism. The effective transition operator has apart
from the terms occuring in the``original'' operator $\widehat{O}$
additional terms that depend on the correlation function $g$.  The
implications of those terms for the two-nucleon knockout reaction have
been illustrated in Fig.~\ref{fig:bosen2}.  In the absence of
correlations that go beyond the IPM only diagrams of the type (a) 
would contribute to a direct two-nucleon emission process.  
Diagram (a)  corresponds with a genuine two-body
absorption process as e.g. encountered in the situation  
in which a photon
is coupled to a charged meson exchanged between two nucleons moving in
a mean-field orbital.  Diagram (d) is of the same form as diagram (a) and
can be interpreted as a short-range correction to the two-body current
contributions from diagram (a).  It
will introduce a decreased probability at short internucleon distances
for two-body photoabsorption processes to occur.  As those short-range
events are already heavily cut by the introduction of hadronic form
factors they have a rather small impact on the calculated cross
sections \cite{vanorden}.

Diagrams (c),(e) and (f) are typical events that are linear in the
correlation function $g$ and involve three and four-body
operators. Given the numerical complexity of two-nucleon knockout
calculations we will restrict ourselves to the two-body operators that
arise from diagrams of the type (a) and (b).  This is often referred to as the
Single-Pair Approximation (SPA).  Within the SPA, one retains those
terms from the above expansion that refer to processes in which the
photon hits two correlated nucleons as a result of which they are
emitted from the target nucleus.  In the calculations we do not
account for reactions in which the photon couples to a nucleon as a
result of which one (or two) nucleons in the recoiling (A-1) system
are emitted.  
Moreover, we did not
consider the final state correlations.  Indeed, within the context of
the SPA inclusion
of the final state correlations would mean that one considers the
correlation between two nucleons residing in a scattering state.  This
type of correlation could be expected to be small.  
To the
best of our knowledge, no calculations are available that address the
correlation function for the continuum states appearing in
the final state of the reactions under study.  Most investigations
into the correlation function have solely addressed the nuclear ground
state.

It is worth stressing that in the absence of ground-state correlations
the one-body photoabsorption operator $\widehat{O}^{[1]}$ would not
contribute to the direct dinucleon emission process. 
  It is only after introducing the central (or Jastrow)
 correlations that it starts
contributing and can be formally treated like e.g. the MEC.  Referring
to Eq.~(\ref{eq:heidi}) it seems technically possible to write down
effective many-body currents that arise from combining one-body
photoabsorption and ground-state correlation effects.
In the SPA, the effective two-body current that accounts for
the short-range corrections in the ground state reads
\begin{equation}
{\vec J}^{[2]}_{SRC}(i,j)= - \left(  {\vec J}^{[1]}(i)
+ {\vec J}^{[1]}(j) \right) g(r_{ij}) \; .
\end{equation}
For the one-body current ${\vec J}^{[1]}$ we consider the standard form
of the Impulse Approximation (IA) that has a convection and
magnetization part.  Consequently, 
\begin{eqnarray}
{\vec J}^{[2]}_{SRC}(i,j) & = & - \frac {e_i} {2iM_N}
\left[ \vec{\nabla} _i \delta({\vec r}-{\vec r}_i) + \delta({\vec r}-{\vec r}_i)
\vec{\nabla} _i \right] g(r_{ij}) \nonumber \\
& & - \frac {e_j} {2iM_N}
\left[ \vec{\nabla} _j \delta({\vec r}-{\vec r}_j) + \delta({\vec r}-{\vec r}_j)
\vec{\nabla} _j \right] g(r_{ij}) \nonumber \\
& & - \frac {\mu _i e} {2M_N} \delta({\vec r}-{\vec r}_i) \vec{\nabla} \times
\vec{\sigma} _i g(r_{ij})
\nonumber \\
& & - \frac {\mu _j e} {2M_N} \delta({\vec r}-{\vec r}_j) \vec{\nabla} \times
\vec{\sigma} _j g(r_{ij}) \; .
\label{srccur}
\end{eqnarray}
Similarly, the effective two-body charge density becomes :
\begin{equation}
\rho ^{[2]} _{SRC}(i,j) = - \left(e_i \delta({\vec r}-{\vec r}_i)
+e_j \delta({\vec r}-{\vec r}_j) \right) g(r_{ij}) \; .
\label{srccha}
\end{equation}
In order to facilitate the calculation of the matrix elements, the
central correlation function is expanded according to :
\begin{equation}
g(r_{12})= \sum _{lm} \frac {4 \pi} {2l+1} g_l(r_1,r_2) Y_{lm} ^{*}(\Omega _1)
Y_{lm} (\Omega _2) \; .
\label{gexpan}
\end{equation}
The different partial-wave components can be directly obtained from :
\begin{equation}
g_l(r_1,r_2) = \frac {2l+1} {2} \int _{-1} ^{+1} dx P_l(x) g 
\left( \sqrt{r_1^2 + r_2 ^2 - 2 r_1 r_2 x} \right) \; .
\end{equation}
The expressions for the two-body transition matrix elements
(\ref{mfel}) and (\ref{mfcoul}) with the effective two-body operators
of Eqs.~(\ref{srccur}) and (\ref{srccha}) are given in Appendix~A.
These matrix elements represent are at the basis of our calculation of the
SRC contribution to exclusive (e,e$'$NN) and semi-exclusive (e,e$'$N)
cross sections.  Results of those calculations will be presented in
the forthcoming sections.  
  In the calculations which will be presented below we have used
several forms of the central correlation function $g$.  Theories
that start from principal grounds, seem to produce correlation
functions that are rather soft.  In this context we refer to the Fermi
Hypernetted Chain (FHNC) calculations of Refs.~\cite{co} and
\cite{co2}, the Monte Carlo calculations
that were e.g. reported in Refs.~\cite{pieper} and \cite{benhar}  and
the recent calculations performed in a translationally invariant
framework \cite{gua96}.  The
object of these studies is usually the ground-state properties of
finite nuclear systems.  Most of these calculations include also
higher order cluster terms.  

To set some kind of upper limit on the effect of the SRC for the 
calculated cross sections, we will also present results with a
pronounced hard-core correlation function that is due to Ohmura,
Morita and Yamada (OMY)
\cite{omy} :
\begin{equation}
f(r) = \left\{ \begin{array}{ll}
0 & r \leq 0.6 fm \\
\left[1-\mathrm{exp}\left(- \mu ^2(r-c)^2 \right) \right] 
\left[1+ \gamma \mathrm{exp} \left( \mu ^2(r-c)^2 \right) \right] & 
r > 0.6 fm
\end{array}
\right. \; ,
\end{equation}
with $ \mu =1.118~fm^{-1}$ and $\gamma$=2.078.
  We do not consider this as a realistic correlation
function but use this function to maximize the effect of the
short-range correlations relative to the meson-exchange and $\Delta$
degrees of freedom.

\section{Results and discussion}
\label{sec:resu}
\subsection{Exclusive (e,e$'$pp) and (e,e$'$pn) reactions}
\begin{figure}
\centering
\epsfysize=18.cm
\epsffile{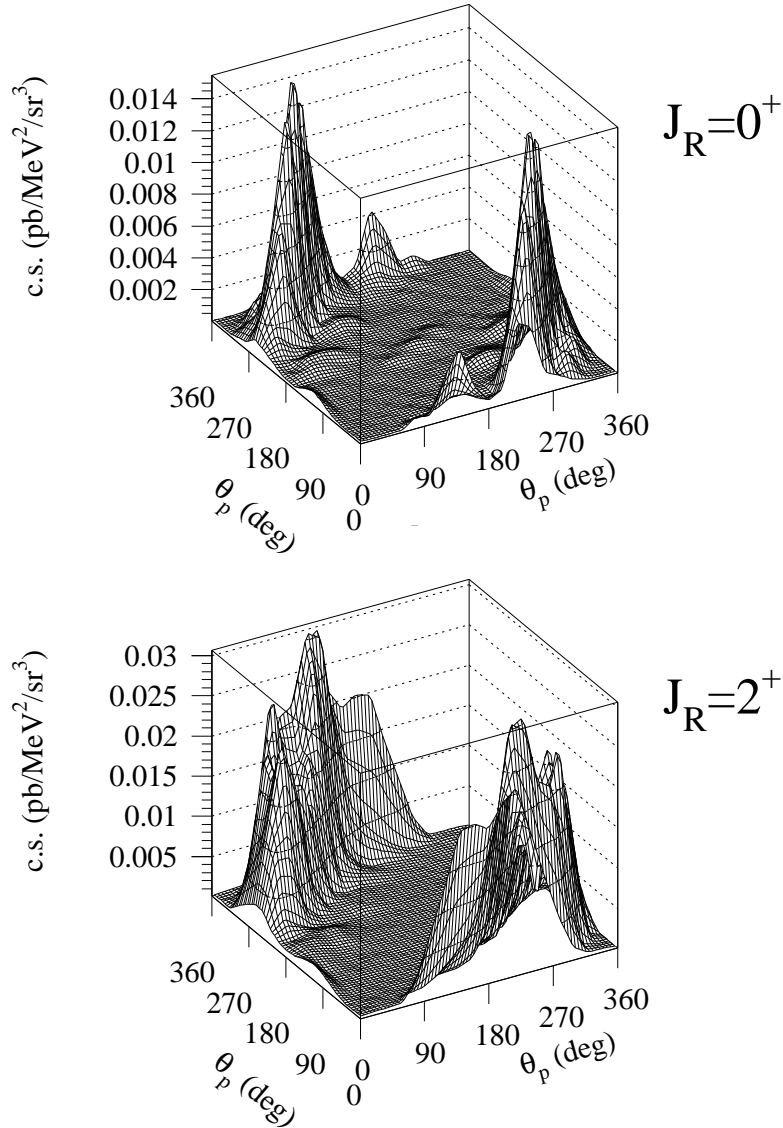}
\caption{Differential cross sections for the $^{12}$C(e,e$'$pp)
reaction at $\epsilon$=700~MeV, $\omega$=250~MeV and $\theta
_e$=30$^o$ (q=383~MeV/c) for excitation of an 
$\mid (1p_{3/2})^{-2} ; 0^+>$ and an 
$\mid (1p_{3/2})^{-2} ; 2^+>$ state respectively.}
\label{fig:d3pp}
\end{figure}

\begin{figure}
\centering
\epsfysize=18.cm
\epsffile{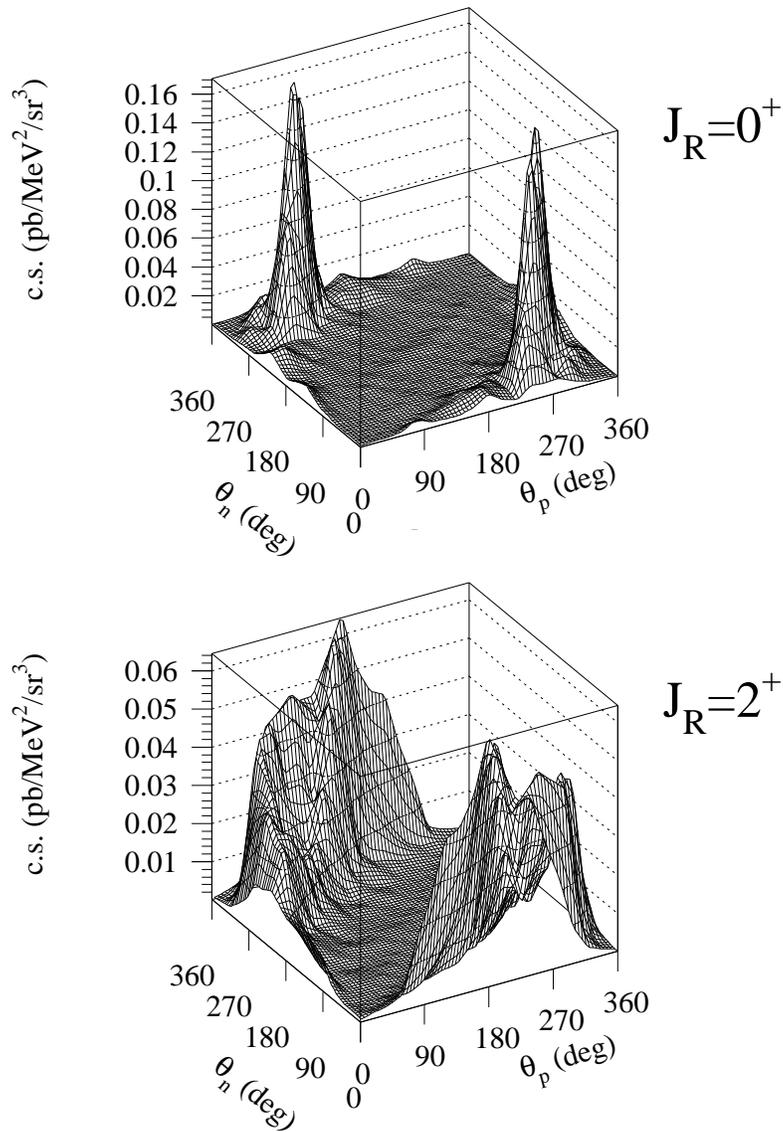}
\caption{As in Fig.~\protect\ref{fig:d3pp} but now for the 
$^{12}$C(e,e$'$pn) reaction}.
\label{fig:d3pn}
\end{figure}

The most elementary (e,e$'$NN) cross sections (c.s.) that can be calculated
are those for excitation of the residual nucleus in a specific
(discrete) state.  Even with fixed the electron kinematics, one is
left with an angular cross section depending on five independent
variables.
An obvious choice for the latter are
the two solid angles and one of
the kinetic energies of the escaping particles :
 $( \Omega _a, \Omega _b, T_{a})$.  
In what
follows we will restrict ourselves to in-plane kinematics which
fixes the azimuthal angle of the two detected nucleons at 0$^{o}$ and
180$^{o}$ degrees.   In Figs.~\ref{fig:d3pp} and  ~\ref{fig:d3pn} 
we show $^{12}$C(e,e$'$pp) and $^{12}$C(e,e$'$pn) angular cross
sections for creation of the final nucleus in a state with
$\mid (1p_{3/2})^{-2}, J_R>$ two-hole structure and $J_R=0^+$ and $2^+$. The
considered kinematical conditions are typical ones for the
800~MeV AMPS (Amsterdam) and MAMI (Mainz) electron accelerators. 
In order
to further reduce the number of independent variables, for the
calculations presented here we
have considered the situation in which both escaping nucleons have
equal kinetic energies at zero recoil momentum $\vec{p}_{A-2}$
\begin{equation}
\vec{p} _{A-2} = \vec{q} - \vec{k}_a -\vec{k}_b \; .
\end{equation}   
This means that the kinetic energy for one of the escaping particles
was fixed at $T=(\omega - S_{2N})/2$, with $S_{2N}$ the threshold
energy for $2N$ emission.
With this choice we are
left with two independent variables, the polar angles of the escaping
nucleons, against which the angular cross sections can be studied. 
In Figs.~\ref{fig:d3pp} and  ~\ref{fig:d3pn} the polar angles vary
between 0$^o$ and 360$^o$.  Within this convention a typical
back-to-back situation in the LAB frame would be e.g. 
($\theta _a = 40 ^o,\theta _b = 220 ^o$). 
The thresholds for
proton-proton and proton-neutron knockout from $^{12}$C are
respectively 27.2 and 27.4~MeV.  Obviously, the two-nucleon
knockout strength is residing in a very small part of the phase space,
namely those situations in which both nucleons
escape in an almost back-to-back situation.  This type of behaviour
can be easily explained by referring to the factorized model for
two-nucleon knockout \cite{Got58,janfac}.  In such an approach it can be
shown that the angular cross sections are proportional to the
c.o.m. distribution $F(P)$ of the active nucleon pair.  As the
 $F(P)$ is a
sharply decreasing function with c.o.m. momentum $\mid \vec{P} \mid$ 
($\vec{P}=-\vec{p}_{A-2}$), two-nucleon
knockout will preferentially occur in those situations in which the
momentum $P$ is kept small.  From energy-momentum conservation
arguments, this corresponds with back-to-back emission.
Despite the fact that in the unfactorized model some of the approximations
which are at the basis of the factorization are not made, we still find
the c.o.m. distribution of the pair to be a driving
mechanism behind two-nucleon knockout.  Besides this back-to-back dominance, 
it is clear that other variables do play a role.  
Inspecting Figs.~~\ref{fig:d3pp} and  ~\ref{fig:d3pn} it is obvious
that 
the angular momentum of the residual nucleus has a
large impact on the shape of the angular cross section.  The angular
distributions for pp and pn knockout leading to a 2$^+$ state are much
wider than those for the corresponding 
0$^+$ state.  Moreover, also the details of the
leading photoabsorption mechanisms reflect themselves in the shapes of
the angular cross sections.  Even for creation of the final nucleus in 
a particular $J_R$ and corresponding kinematical conditions, 
the proton-proton
and proton-neutron cross section show some differences which can be
related to the reaction 
mechanisms contributing to the cross sections.  In this
context, we want to remind the reader of the fact that the
meson-exchange (Eq.~(\ref{eq:pisea})) and isobaric 
(Eq.~(\ref{eq:pidelta})) 
currents have a peculiar isospin
dependency which makes them behaving differently in the proton-proton
and proton-neutron channel.        

\begin{figure}
\centering
\epsfysize=18.cm
\epsffile{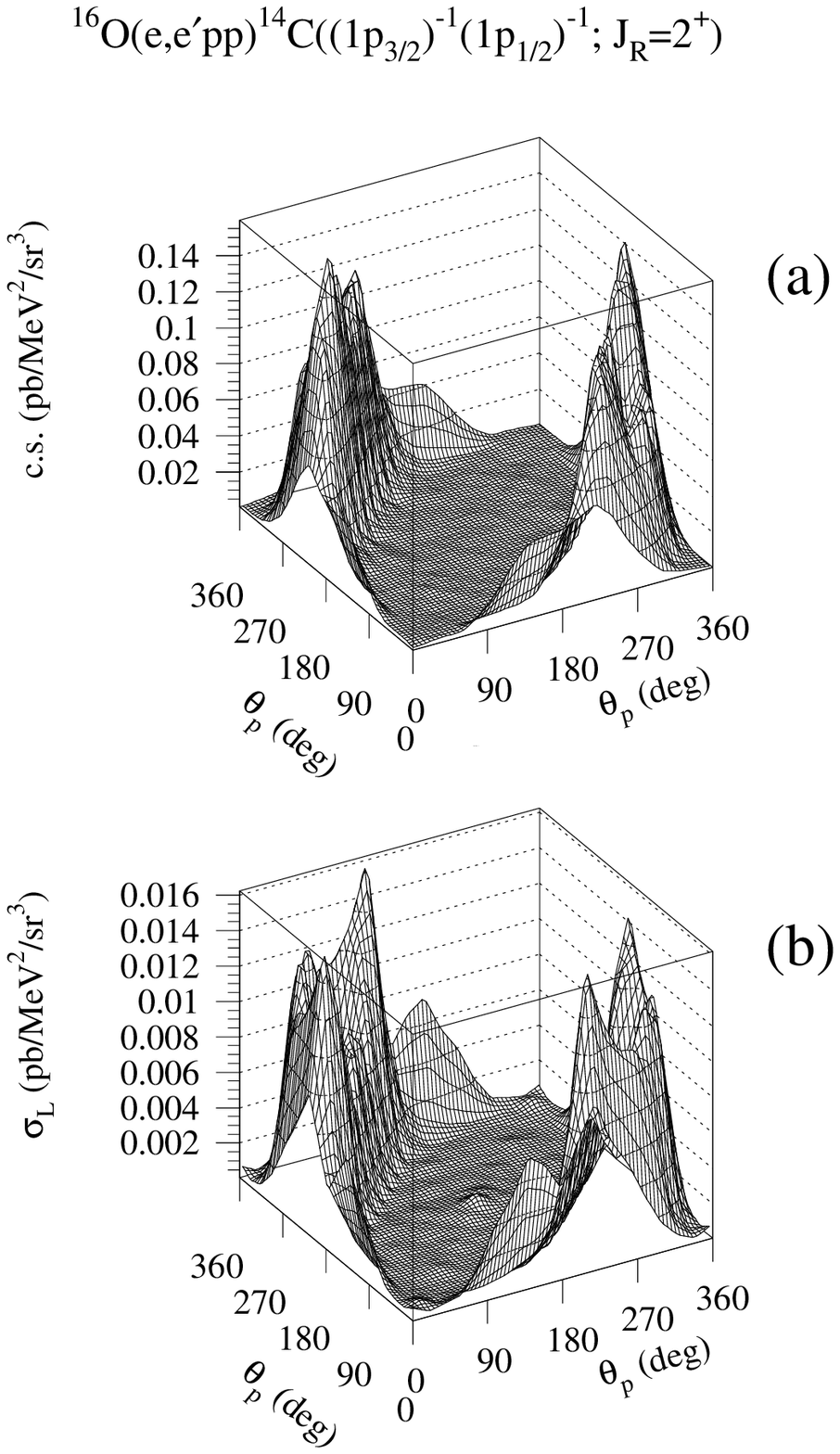}
\caption{Differential cross sections for the $^{16}$O(e,e$'$pp)
reaction at $\epsilon$=1.2~GeV, $\omega$=225~MeV and $\theta
_e$=12$^o$ (q=319~MeV/c) for excitation of an 
$\mid (1p_{3/2})^{-1}(1p_{1/2})^{-1} ; J_R=2^+>$ two-hole state. (a) full cross
section (b) longitudinal contribution.}
\label{fig:pp225}
\end{figure}

\begin{figure}
\centering
\epsfysize=18.cm
\epsffile{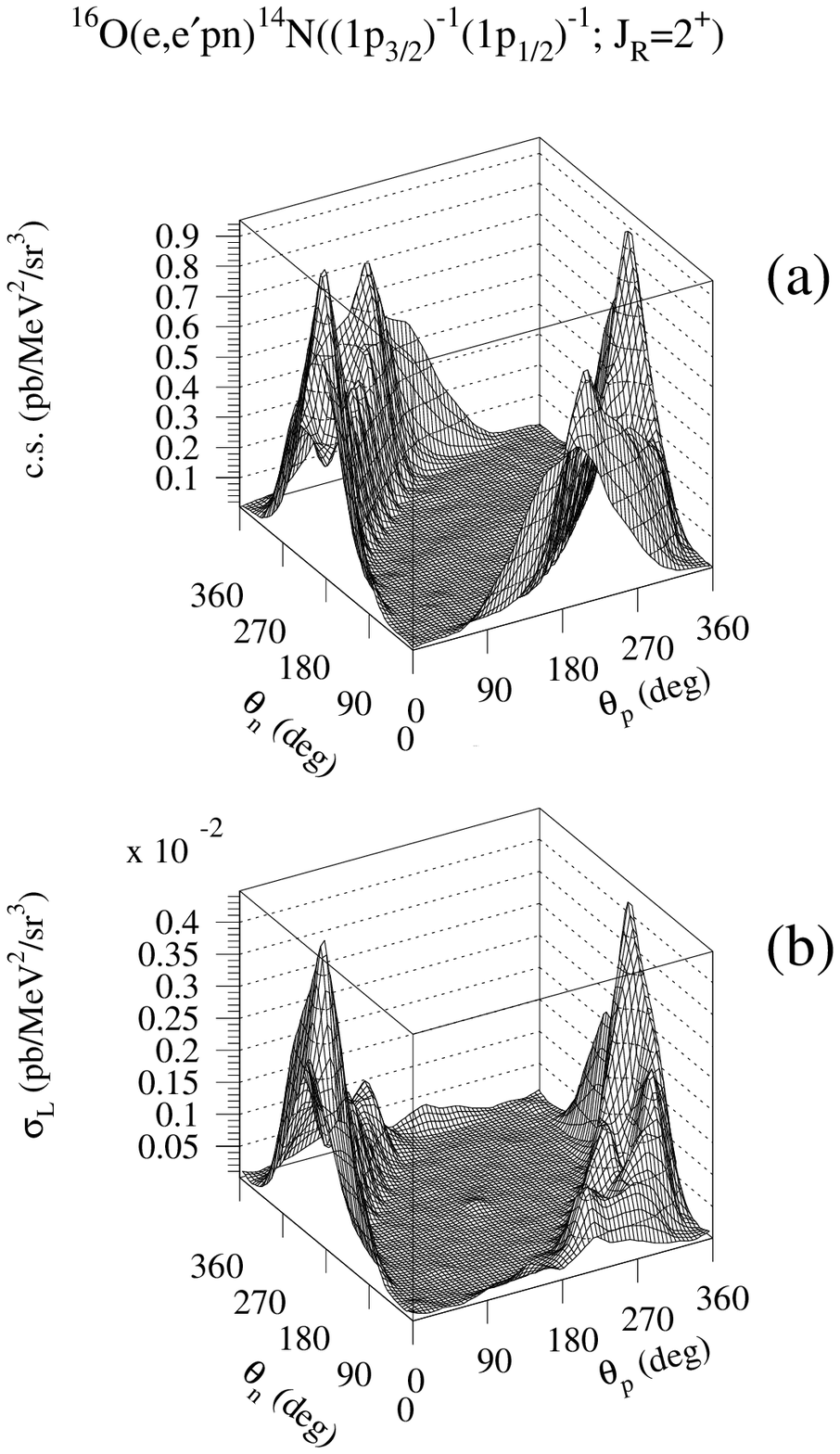}
\caption{As in Fig.~\protect \ref{fig:pp225} but now for
proton-neutron emission.}
\label{fig:pn225}
\end{figure}

\begin{figure}
\centering
\epsfysize=18.cm
\epsffile{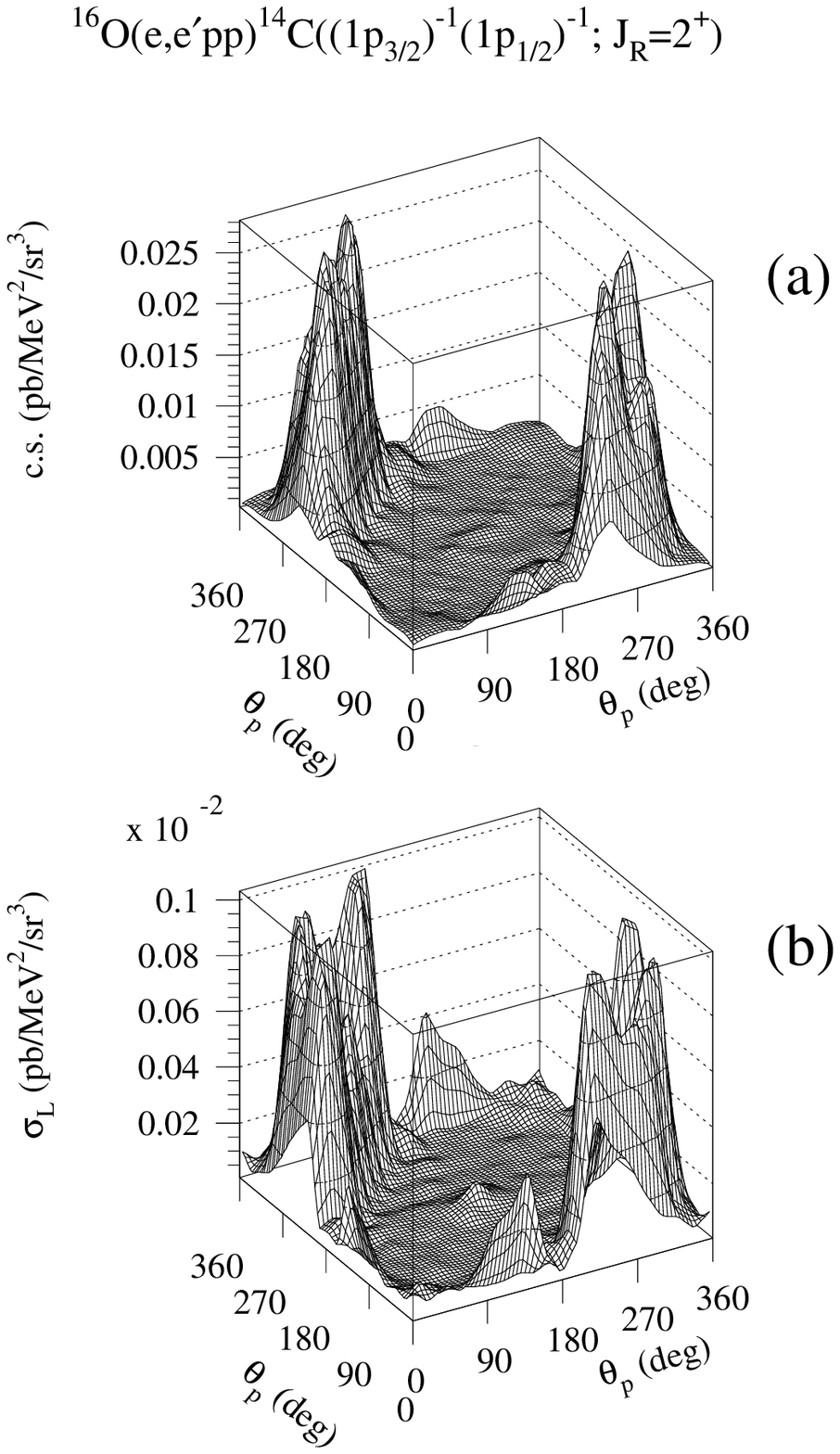}
\caption{Differential cross sections for the $^{16}$O(e,e$'$pp)
reaction at $\epsilon$=1.2~GeV, $\omega$=375~MeV and $\theta
_e$=12$^o$ (q=429~MeV/c) for the excitation of an 
$\mid (1p_{3/2})^{-1}(1p_{1/2})^{-1} ; J_R=2^+>$ two-hole state. (a) full cross
section (b) longitudinal contribution. }
\label{fig:pp375}
\end{figure}

\begin{figure}
\centering
\epsfysize=18.cm
\epsffile{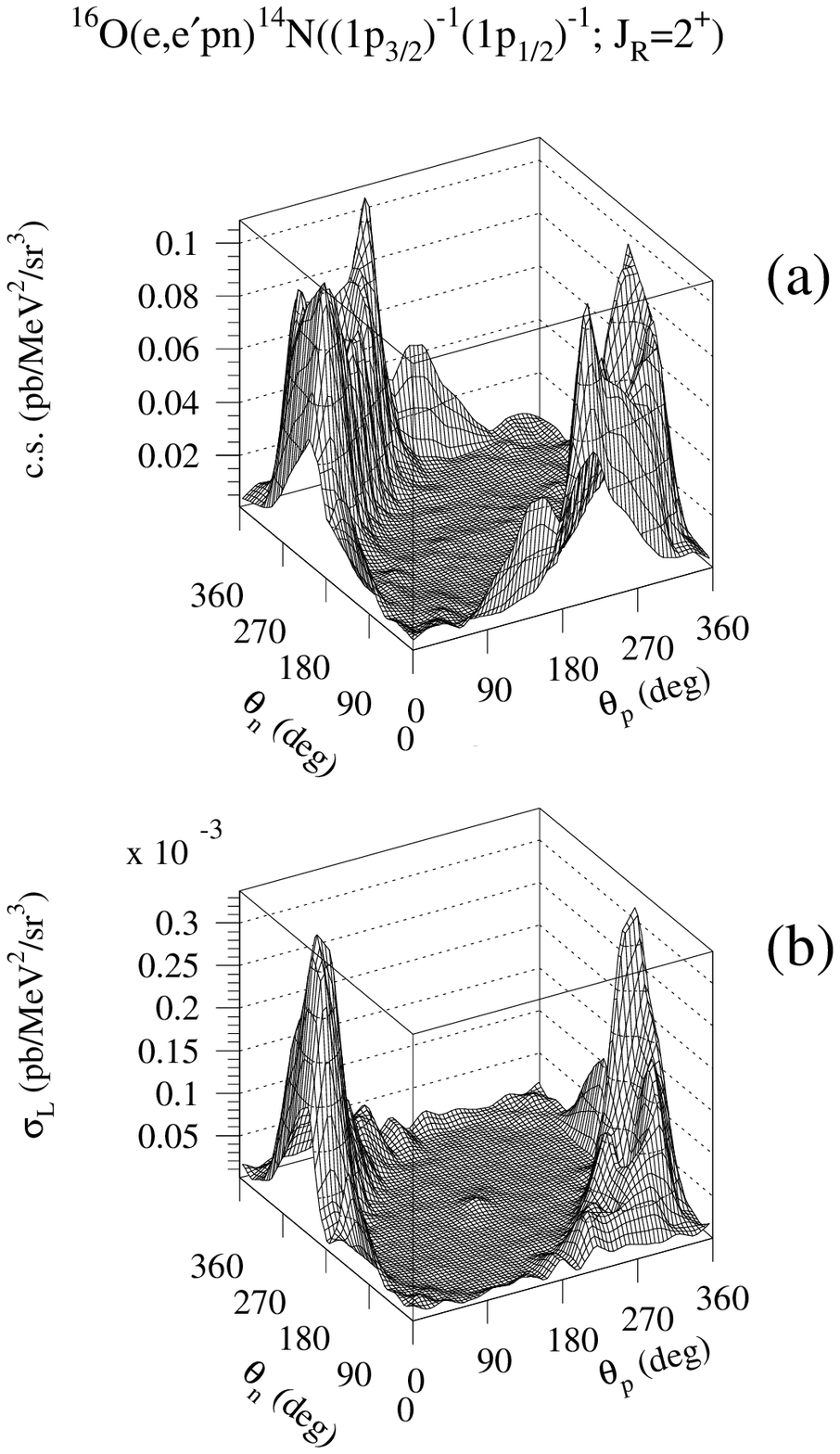}
\caption{As in Fig.~\protect \ref{fig:pp375} but now for
proton-neutron emission.}
\label{fig:pn375}
\end{figure}

One of the main tasks of this paper is to investigate the sensitivity
of the (e,e$'$NN) cross sections to dynamical short-range
correlations.   For that purpose we have calculated 
$^{16}$O(e,e$'$pp) and (e,e$'$pn) 
angular cross section for emission out of the
p-shell orbitals.    A measure of the relative importance of the
ground-state correlations in the whole reaction process, 
is the contribution from the longitudinal
channel to the differential cross sections.  Indeed, into first order 
the longitudinal
 channel is free from
isobaric and
meson-exchange current contributions. In the presented model
calculations the longitudinal strength is 
exclusively the result of the existence of dynamical
short-range correlations.  A measure for the degree of longitudinal
polarization of the virtual photon is the parameter $\epsilon _L$
\begin{eqnarray}
2 \epsilon _L  & \equiv & \frac {v_C} {v_T} \nonumber \\
& & = \left( tg^2 \frac {\theta_e} {2} - \frac {q_\mu q^\mu} 
{2 \mid \vec q \mid ^2} \right) ^{-1} \; .
\label{eq:epsl}
\end{eqnarray}
In
Figs.~\ref{fig:pp225}-\ref{fig:pn375} we have plotted some
corresponding proton-proton and proton-neutron knockout angular 
cross sections
at two values of the energy transfer.  These have been chosen as to
probe a kinematical regime well below and above the $\Delta$ resonance
region (the equivalent photon energies $K$ are 198 and 352~MeV
respectively).   The large incident electron energy
($\epsilon$=1.2~GeV) and small electron scattering angle ($\theta
_e$=12~$^o$) considered make the longitudinal polarization $\epsilon
_L$ of the virtual photon large.  For the results of 
Figs.~\ref{fig:pp225}-\ref{fig:pn375}, we have considered the
situation in which the residual nuclei 
$^{14}$C and $^{14}$N are
created in a 
$\mid (1p_{3/2})^{-1}(1p_{1/2})^{-1} ; J_R=2^+>$ state.  According to
the nuclear-structure calculations of Ref.~\cite{geurts} this type of
configuration 
would be preferentially populated in $^{16}$O(e,e$'$pp) reactions.
Furthermore, we have considered
the situation in which both escaping  nucleons have the same kinetic
energy at $\mid \vec{p}_{A-2} \mid =0$.  The proton-neutron channel is
obviously a much stronger channel than the proton-proton emission
channel.  Inspecting Figs.~\ref{fig:pn225} and \ref{fig:pn375} it is
clear that proton-neutron emission is dominated by the transverse
channel and that central 
ground-state correlations play only a marginal role.  

The weaker proton-proton emission channel
exhibits a stronger sensitivity to the ground-state correlations.
Over the whole, however, the isobaric currents dominate the angular
cross sections.  There are some parts of the phase space, however,
for which the longitudinal cross section is a sizeable contribution
in the full cross section, reflecting the sensitivity of the
reaction to ground-state correlation effects.  The latter are e.g.
playing a major role for kinematical conditions that correspond with
one of the escaping nucleons moving along the direction of the
three-momentum transfer $\vec q$.         

All of the previous angular (e,e$'$NN) 
cross sections have been obtained with the
central FHNC correlation function.  The latter is of the Gaussian type
$g(r_{12})=\alpha e ^{-\beta r_{12}^2}$.  For $^{16}$O, the parameters
were calculated to be $\alpha$=0.53 and $\beta$=1.52~fm$^{-2}$
\cite{co}.  For our purposes, 
the parametrization for $^{12}$C was assumed to be
identical to the one for $^{16}O$. 

\begin{figure}
\centering
\epsfysize=9.cm
\epsffile{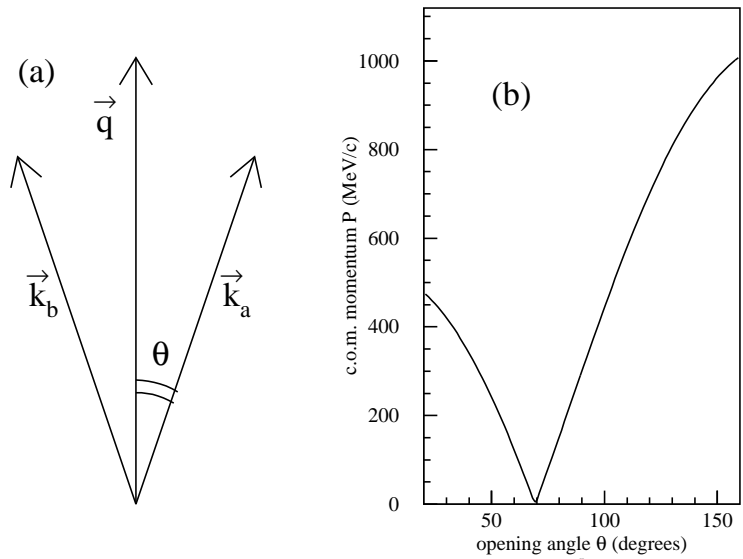}
\caption{(a) Coplanar and symmetrical kinematics. (b) the c.o.m.
momentum $P$ as a function of the opening angle $\theta$ for
two-proton knockout from $^{16}$O at an energy transfer of $\omega$=210~MeV.}
\label{fig:cofp}
\end{figure}

\begin{figure}
\centering
\epsfysize=11.cm
\epsffile{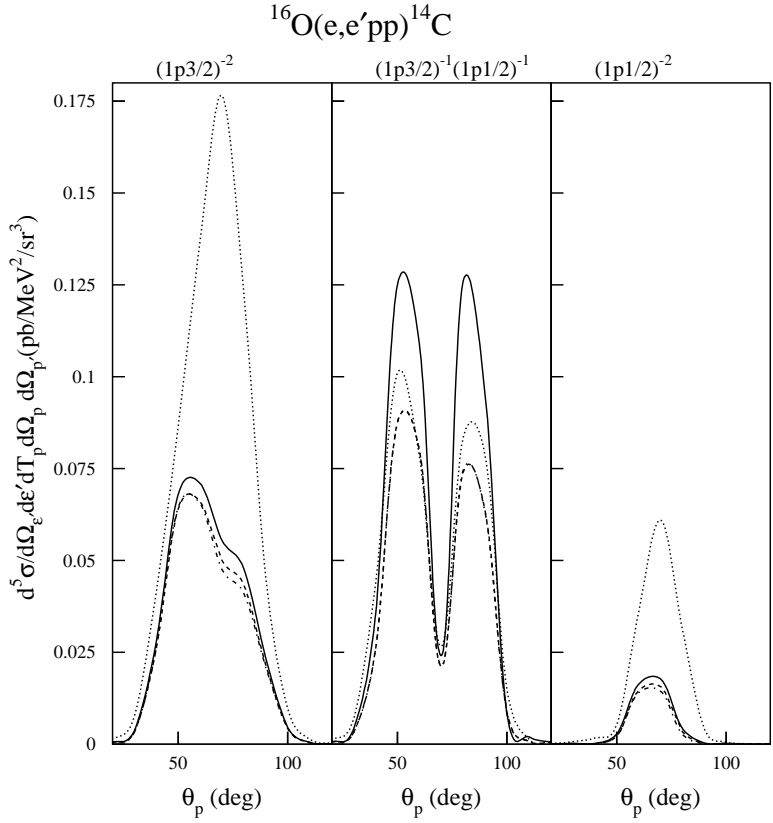}
\caption{The $^{16}$O(e,e$'$pp) differential cross section as a
function of the proton angle in coplanar
and symmetrical kinematics for $\epsilon$=516~MeV, $\omega$=210~MeV
and $\theta _e$=31$^o$.  The different combinations for emission out
of the p-shell orbits are considered.  
Different central correlation functions have been used : OMY
(dotted line), FHNC (solid line) and MC (dashed line).  The dot-dashed
line is the result of a calculation in which only the isobaric
currents are included.}
\label{fig:cross}
\end{figure}

We now aim at studying the sensitivity of
the calculated (e,e$'$pp) cross sections to the choice of the Jastrow
correlation function.  To this end,
we do not longer consider the
situation in which the residual nucleus is created in a particular
$J_R$ but plot the cross section which is obtained after
(incoherently) adding all possible final angular momenta $J_R$.  The
number of situations which have to be considered depends on the
quantum numbers ($l_h j_h, l_{h'} j_{h'}$)
of the orbits from which the nucleons are escaping.  For example, for
proton-proton emission out of the $(1p_{3/2},1p_{1/2})$
orbits the plotted cross section is obtained after adding the
contribution from $J_R=1^+$ and $2^+$.  For the results of
Fig.~\ref{fig:cross} we consider the so-called coplanar and
symmetrical situation.  
This type of
 kinematics has been schematically sketched in Fig.~\ref{fig:cofp}  
and corresponds with the situation in which both nucleons are escaping
with equal kinetic energy and opening angle with respect to the
direction of the transferred momentum.  In  Fig.~\ref{fig:cofp} we
show also the typical behaviour of the c.o.m. momentum $P$ in such
kinematics. So given the dominance of the c.o.m. distribution in
producing the general features of the two-nucleon knockout cross
sections, one
would expect the angular cross section to peak around an opening angle
of 70 degrees and fall off quickly to smaller and larger opening
angles.  This is precisely what is observed in Fig.~\ref{fig:cross}.   
Another observation is that with the central correlation function of
Ref.~\cite{pieper}, which is obtained with Monte-Carlo techniques, 
the Jastrow correlations are hardly affecting the differential
cross sections that are obtained when solely accounting for the
$\Delta$ degrees of freedom.  In the remainder of the paper, the
correlation function of Ref.~\cite{pieper} will be referred to as
``MC''.    
A more pronounced effect is found with the Gaussian
FHNC correlation function.  The short-range contribution is further
noticed to depend
on the single-particle levels from which the protons are escaping.
Whereas the Gaussian FHNC
correlation function slightly increases the cross sections for
$(1p1/2)^{-2}$ and $(1p3/2)^{-2}$ knockout, the effect is quite sizeable for
2N knockout out of the mixed $(1p1/2)^{-1}(1p3/2)^{-1}$ orbits.  
The most spectacular shell dependence is
noticed for the hard-core OMY correlation function.  Counting the
number of possible proton-proton pairs one would expect the absolute
cross sections for the different two-hole configurations
 of Fig.~\ref{fig:cross} to be like 
6~($(1p3/2)^{-2}$):8~($(1p3/2)^{-1}(1p1/2)^{-1}$):1($(1p1/2)^{-2}$).
It is obvious that the number of pairs can only serve as a rough guide
to predict the relative amount of strength that goes into the
different two-hole combinations.

\begin{figure}
\centering
\epsfysize=11.cm
\epsffile{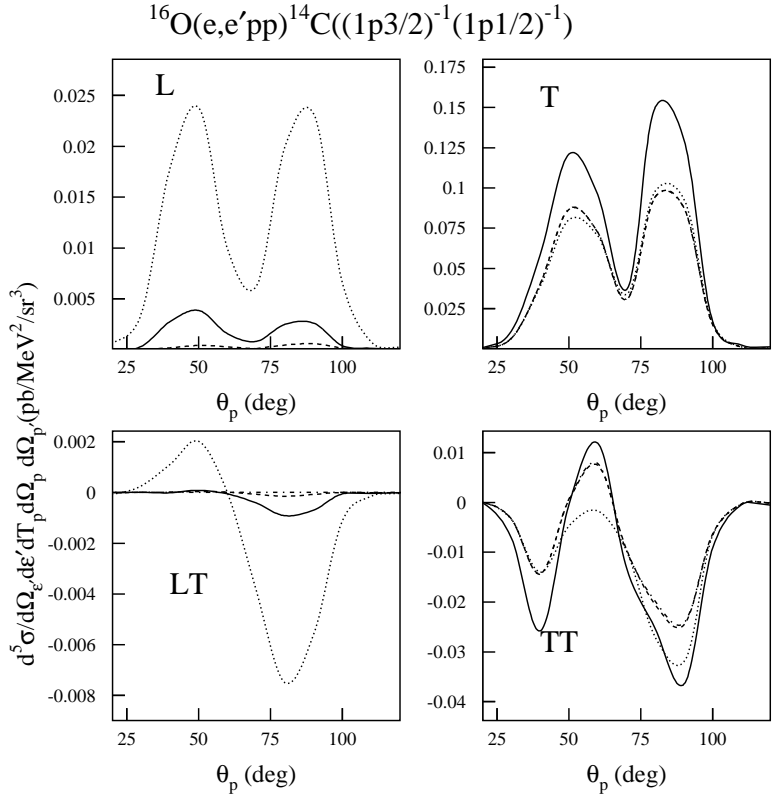}
\caption{The four terms contributing
to the $^{16}$O(e,e$'$pp)((1p3/2)$^{-1}$(1p1/2)$^{-1}$)
angular cross sections 
plotted in the
central panel of Fig.~\protect \ref{fig:cross}.  The dot-dashed line
shows the result when including only the delta-currents.  The
other curves are obtained after adding also the central short-range
effects.  Different central correlation functions have been used : OMY
(dotted line), FHNC (solid line) and MC (dashed line).}
\label{fig:stru23}
\end{figure}

In Fig.~\ref{fig:stru23} the four terms (longitudinal, transverse,
longitudinal-transverse and transverse-transverse)
contributing to the
$^{16}$O(e,e$'$pp)((1p3/2)$^{-1}$(1p1/2)$^{-1}$)
angular cross sections of Fig.~\ref{fig:cross} are shown for 
different choices of the central
correlation function.  It is clear that the absolute magnitude of the
cross section is dominated by the transverse (T) and
transverse-tranverse (TT) terms that show very little sensitivity to
the central correlation effects.  The longitudinal (L) and
longitudinal-transverse (LT) structure functions, on the other hand,
exhibit a very strong sensitivity to the choice of the correlation
function.  The hard-core OMY correlation function produces
longitudinal (e,e$'$pp) strength that is typically one order of
magnitude larger than the strength generated by the (realistic)
soft-core 
MC and
FHNC correlation functions.  Remark further that the MC and FHNC
correlation functions produce remarkably different L and LT structure
functions.  This is a rather surprising result as their functional
dependence on the relative coordinate of the two interacting particles
is rather similar.

\subsection{Semi-exclusive (e,e$'$p)}

In the plane-wave impulse approximation (PWIA) the (e,e$'$p) cross
section can be cast in the form 
\begin{equation}
\frac {d^4 \sigma} {d \epsilon ' d \Omega _{e'} d \Omega _{p} dE _p}
=  E p  \sigma _{ep} P( \left| \vec{p}_m \right|, E_m) \; ,
\label{eq:factor}
\end{equation}
where E$_p$ (p$_p$) is the energy (momentum) of the detected proton and
$\sigma _{ep}$ the elementary cross section for electron scattering off
an off-shell proton.  It should be noted that a similar type of
factorization can be pursued for the ($\gamma$,p) reaction.
The spectral function $P( \left| \vec{p}_m
\right|, E_m)$ is related to the probability of removing a nucleon
with momentum $p_m$ from the target nucleus and finding the residual
nucleus at a missing  energy $E_m$.  The spectral function  is
of fundamental importance for the understanding of the many-body
dynamics of finite nuclei and has been the subject of many theoretical
considerations \cite{hodgson,sick94,benh94}.  
Eventhough a wide variety of spectral functions is
available in literature, most calculations point towards the IPM
providing a fair description of $P(k,E)$ at low $k$ and $E$.  As far
as the spectral function is concerned, the deviations from the IPM
model are predicted to occur at high energy and momentum.  As a
consequence, when trying to access these correlation effects
 with the aid of
(e,e$'$p) measurements, one has to probe high missing energies, where
the exclusive nature of the reaction 
can no longer be guaranteed.  In this missing-energy range, 
the usefulness of the
factorized cross section (\ref{eq:factor}) becomes questionable.
Indeed, contributions from genuine two-body photoabsorption, like e.g.
those mediated by pion exchange, to the (e,e$'$p) cross section cannot
be related to the single-particle spectral function, even when
adopting a plane wave description for the escaping particles.  On the
other hand, one-body photoabsorption with subsequent emission of two
nucleons is a signature of ground-state correlations and has also been
predicted to be the major mechanism \cite{ciofi96} feeding the high-momentum
components in $P(k,E)$.  Or put in other words, just as for the
(e,e$'$NN) reaction, meson-exchange and isobaric currents are
considered as unwanted noise 
 when embarking on studies that aim at probing
ground-state correlations with the semi-exclusive (e,e$'$p) and
($\gamma$,p) reaction.   

\begin{figure}
\centering
\epsfysize=16.cm
\epsffile{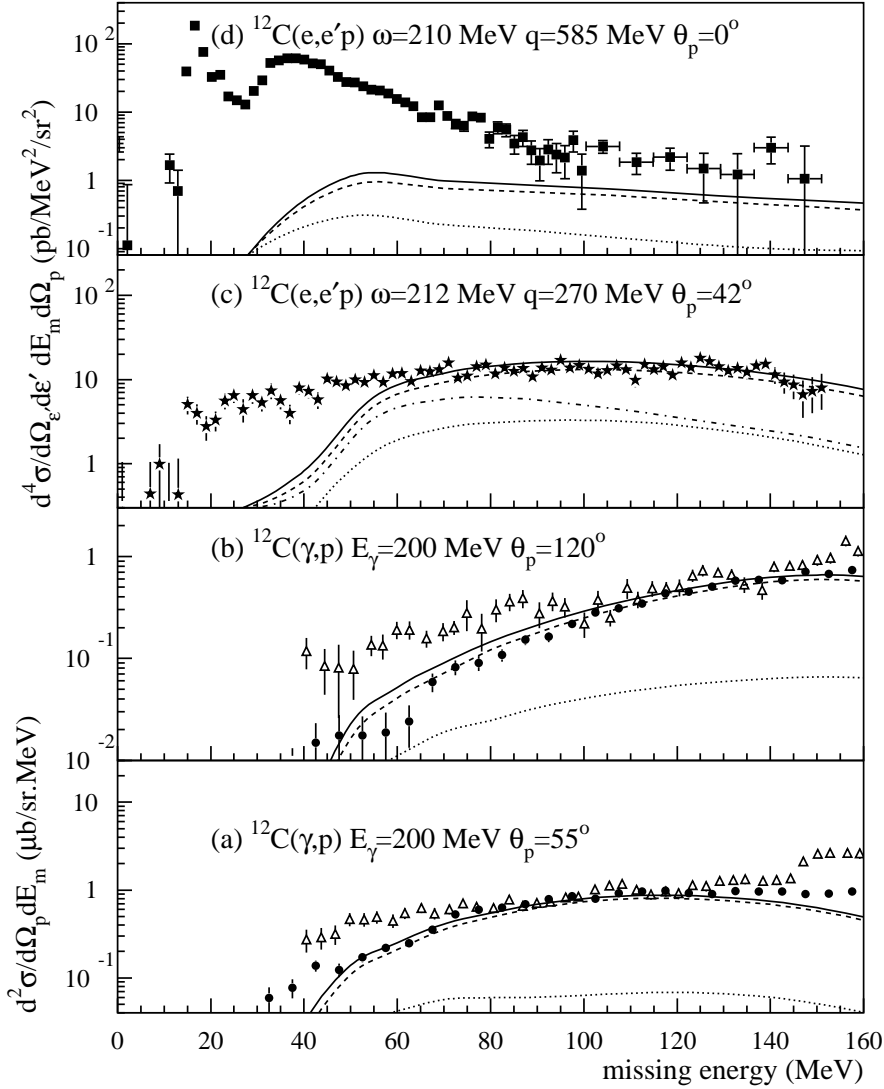}
\caption{Missing-energy dependence of the semi-exclusive
$^{12}$C(e,e$'$p) and $^{12}$C($\gamma$,p) spectrum for $\omega
\approx$ 200~MeV and various kinematical conditions.  The dotted
(dashed) line
shows the calculated contribution from proton-proton (proton-neutron)
emission. 
The
solid line is the incoherent sum of all the calculated strength
contributions.
The dot-dashed line is the calculated contribution from
two-nucleon knockout including only the transverse two-body currents
(for most cases this line falls on top of the solid one).
   The data are from Refs.~\protect \cite{cross} (circles),
\protect \cite{Wei90} (squares),
\protect \cite{anghi} (triangles) and
 \protect \cite{leonplb} (stars)
.}
\label{fig:compar}
\end{figure}

In order to study the q-dependence of the semi-exclusive proton
knockout channel we
have gathered in Fig.~\ref{fig:compar} a number of 
 $^{12}$C spectra that have
been obtained for an energy transfer of about 200 MeV.  The data
include both real and virtual photon results and are obtained at
several labs.  A striking feature of the data is the ratio of the
continuum strength (E$_m \geq $ 30~MeV) to the strength residing in
the discrete part of the spectrum.  For the real photon case the
continuum strength dominates the spectrum.  For the highest momentum
transfer considered here (q=585 MeV) the strength for one-nucleon 
removal from the
1p and 1s shell is clearly showing up.  The real photon data do not
show any obvious indication for one-proton emission from the 1s-shell 
and the strength rises clearly beyond 
the two-particle emission threshold.  An
interesting case is the virtual photon spectrum that was obtained at
low momentum transfer (q=270~MeV).  Here the strength is a rather flat
distribution that extends into the region of s-shell removal without
reflecting a clear bump.  Also shown in Fig.~\ref{fig:compar} is the
calculated contribution from proton-proton and proton-neutron emission
to the semi-exclusive spectrum.  These calculations have been
performed according to the formalism outlined in Sect.\ref{sec:semi}
and include the meson-exchange and isobaric currents in addition to
the strength generated by the Jastrow correlations.  The latter are
accounted for in the FHNC parametrization.
 Apart from the highest momentum
transfer considered here, the calculations give a reasonable account of
the measured strength.  The major contribution to the semi-exclusive
spectrum is ascribed to proton-neutron knockout.  This channel is
predominantly fed by the meson-exchange and isobaric currents,
which imposes some serious limitations on the suitability of the 
semi-exclusive reaction to probe ground-state correlations.    
Referring to Fig.~\ref{fig:compar},
it should be noted that the data taken at
q=270~MeV are probing the dip region, whereas at q=585~MeV one is
facing quasi-elastic conditions.  One could thus conclude that
semi-exclusive $(\gamma^{(*)},p)$ processes are reasonably well 
understood at the
real photon point and in the dip region.  The situation is however
quite different in the quasi-elastic regime for which the origin of the
continuum strength cannot be explained in terms of two-nucleon
knockout. 

\begin{figure}
\centering
\epsfysize=10.cm
\epsffile{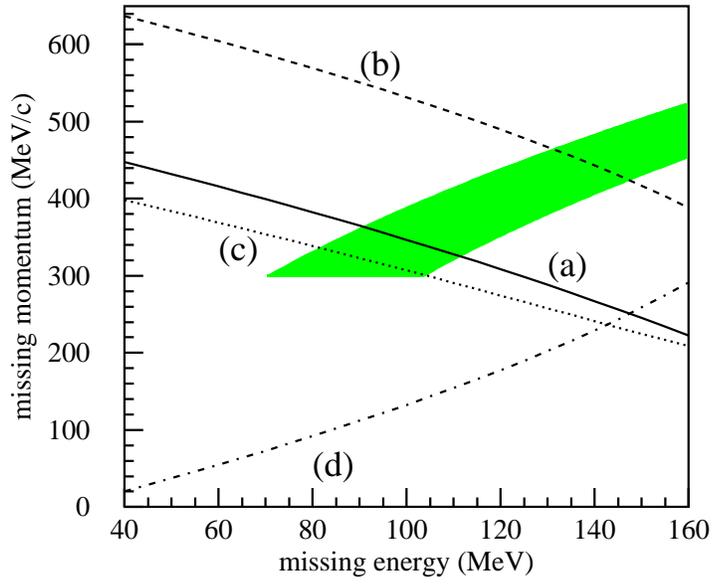}
\caption{The missing momentum as a function of the missing energy for
the kinematical conditions of Fig.~\protect \ref{fig:compar}.}
\label{fig:empm}
\end{figure}

For all four situations considered in Fig.~\ref{fig:compar} only
situation (c)
exhibits some sensitivity to the Jastrow correlations. 
For the other three cases photoabsorption on the meson-exchange and
$\Delta$ currents dominates the calculated two-nucleon knockout
strength.  Under the kinematical conditions of (c) data are available
for proton escaping angles varying from 27~$^o$ up to 162~$^o$. In
Ref.~\cite{janold}, we have compared our (e,e$'$p) model predictions
for the full range of proton angles.  A selective sensitivity to
dynamical SRC was observed.  The qualitative behaviour of this
sensitivity could be explained within the context of the two-nucleon
correlation model \cite{ciofi96} that predicts an increased
sensitivity to SRC effects when the following relation between the
missing energy and momentum is (approximately) obeyed       
\begin{equation}
E_m \approx S_{2N} + \left< E_x^{hh'} \right> + 
\frac {(A-2) p_m^2} {2(A-1)M_N} \; ,
\label{eq:ciofi}
\end{equation}
where $\left< E_x^{hh'} \right>$ is the average excitation energy of
the $A-2$ system if the nucleons are escaping from the orbits
characterized by $h$ and $h'$.  Apart from some factors that are
related to recoil effects, the above relation reflects a picture in
which correlated nucleons occur in pairs with respective momenta
$\vec{p}_m$ and $-\vec{p}_m$.  The kinematical conditions of
the four situations considered in 
Fig.~\ref{fig:compar} are shown in an $(E_m,p_m)$ graph in
Fig.~\ref{fig:empm}. The shaded region is the area for which the above 
prescription (\ref{eq:ciofi}) would predict an increased likelihood to
detect ground-state correlation effects.  The width of the shaded
region  was
obtained by assuming that $0 \leq \left< E_x^{hh'} \right> \leq
50~MeV$, thus covering knockout from the (1p)$^2$, (1p)(1s) and
(1s)$^2$ orbitals in $^{12}C$.  It was further assumed that
correlation effects only occur for $p_m \geq 300~MeV/c$.   
Despite the fact that some likelihood to observe SRC
effects is predicted for the real-photon cases (a) and (b), the transverse
character of the reaction makes the two-body currents dominating.  For
the situation (c) a proper balance between the longitudinal/transverse
degrees of freedom and the condition (\ref{eq:ciofi}) seems to be
reached.          
 
\begin{figure}
\centering
\epsfysize=12.cm
\epsffile{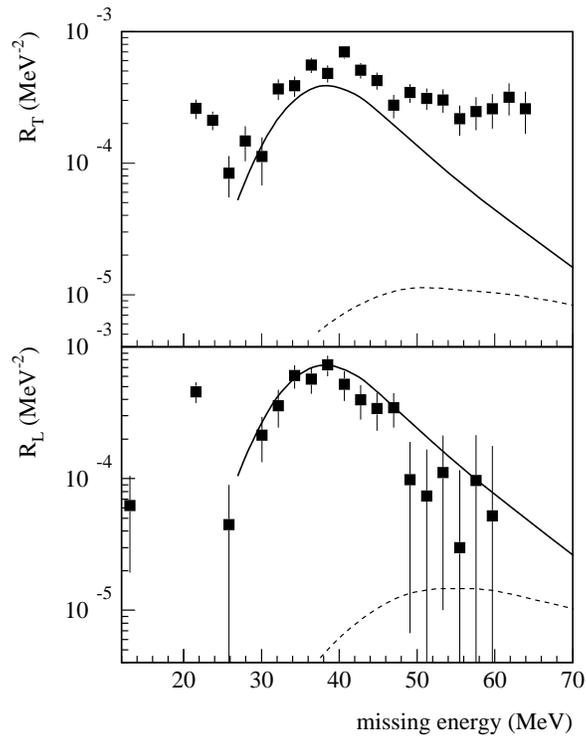}
\caption{The longitudinal and transverse structure functions versus
missing energy for the $^{12}$C(e,e$'$p) reaction at $\omega$=122.5~MeV and
q=397~MeV/c. The dashed line is the calculated contribution from
proton-proton and proton-neutron emission.  The solid line is the
predicted contribution for one-proton knockout from the 1s shell
applying the same spectroscopic factor in the longitudinal and
transverse structure function.  The
data are from Ref.~\protect \cite{Ulm87}.}
\label{fig:ulmer}
\end{figure}

A possible way of learning more about the short-range correlations 
is the
separation of the different (e,e$'$p) structure functions.  In
particular, the longitudinal response function
opens good perspectives in that respect.  As explained in
Sect.~\ref{sec:mec}, in lowest relativistic order
the pion exchange effects are not affecting the longitudinal channel. 
For finite nuclei, only one experiment that separates the different
structure functions in the high missing-energy region has been reported.
In Ref.~\cite{Ulm87} the $^{12}$C(e,e$'$p) structure functions have been
separated in parallel kinematics (${\vec p}_p \parallel {\vec q}$). 
Parallel kinematics is quite favourable for these purposes as only two
structure functions ($W_T$ and $W_L$) are contributing.  The data of
Ref.~\cite{Ulm87} are shown in Fig.~\ref{fig:ulmer} and are
characterized by an excess strength in the transverse response function
in the missing energy region above the s-shell peak. The transverse ($R_T$)
and longitudinal ($R_L$) structure functions of Fig.~\ref{fig:ulmer}
are obtained by dividing the longitudinal (transverse) part of the
(e,e$'$p) cross section by $\sigma _M v_C$ ($\sigma _M v_T$).  
We have investigated in how far the observed excess strength in the
transverse channel can be attributed
to photoabsorption on correlated nucleon pairs with emission of two
nucleons. In these calculations we have included the absorption
mechanisms related to the SRC, IC and MEC.  As becomes clear from
Fig.~\ref{fig:ulmer}, the present model cannot account for the excess
transverse strength.  Remark that in the presented model calculations
all two-nucleon emission strength in the longitudinal channel is
attributed to short-range effects.   

As we find that two-nucleon
knockout cannot explain the excess transverse strength at high missing
energies, we have investigated in how far single-nucleon knockout from
the deep-lying shells could be at the origin of the different
missing-energy behaviour in the longitudinal and transverse channel.
For that purpose we have calculated the cross section for knockout
from the 1s$_{1/2}$ orbit at various values of $T_p$, covering the whole
missing-energy range of the data.  These calculations were done in a
direct knockout model including both one- and two-body currents in the
initial photoabsorption process \cite{highpm}.  For the whole missing-energy range
covered here ($E_m \leq$ 70~MeV) we use the bound state 1s$_{1/2}$ wave
function as obtained from a HF calculation.
Formally, this means that in the calculations the spectral function 
\begin{eqnarray}
P_{lj}(k,E) & = & \sum _{n} \mid \left< \psi _n ^{A-1} (E_n^{A-1}) 
\mid c_{ljk} \mid
\psi _o (E_o^A \right> \mid 
^2 
\nonumber \\
& & \times \delta \left( E - \left(E_n^{A-1}-E_o^A
\right) \right) \; ,
\end{eqnarray}
is used in a factorized form :
\begin{equation}
P_{lj}(k,E) = \mid \phi _{lj} (k) \mid ^2 S_{lj}(E) (2j+1) \; .
\label{eq:appro}
\end{equation}
The function $S_{lj}$ describes the spreading of the hole strength
characterized by the quantum numbers $lj$ as
a function of the missing energy in the A-1 system  
\begin{equation}
S_{lj}(E)= \sum _{n} \left| \left< \psi _n ^{A-1} (E_N^{A-1})
\left| c_{lj} \right|
\psi _o (E_o^A) \right> \right| ^2 \delta \left( E - \left( E_n^{A-1} -
E_o^A \right) \right) \;,
\label{eq:factp}
\end{equation}
and $\phi _{lj}(k)$
is the mean-field single-particle wave function in momentum space.  
Microscopic calculations of the single-particle spectral function 
\cite{mutter}, however, show that deviations from mean-field
wavefunctions predominantly occur at higher missing energies ($E_m$
$ \geq $ 50~MeV).
As such, the factorized prescription (\ref{eq:factp}) could be considered as 
 a reasonable approximation as long as relatively low missing energies
are probed.  In
order to account for the spreading of the 1s hole strength in $^{11}$B
we rely on a prescription which is due to
Jeukenne and Mahaux
\cite{Jeu83}.  In this parametrization the nucleon hole spectral
function is determined by a Lorentzian with an energy-dependent
width $\Gamma (E)$ :
\begin{equation}
S_{lj}(E) = \frac {1} {2 \pi} 
\frac {Z_{lj} \Gamma (E)} { \left( 
E - \left| \epsilon _{lj} 
\right| \right)^2 + \frac {1} {4} \left( \Gamma (E)  \right)^2} \;,
\end{equation}  
where $Z_{lj}$ ($\epsilon_{lj}$) is the quasi-particle strength 
(energy) of the hole state under
consideration. 
The width $\Gamma (E)$ is then related to the imaginary part $W$ of
the optical potential $( \Gamma = 2 W)$ for which the following
parametrization is adopted 
\begin{equation}
W(E)= \frac {9. (E-S_N)^4} {(E-S_N)^4+(13.27)^4} \;\; \mathrm{(MeV)} \; . 
\end{equation}
With the above assumptions a fair description of the missing-energy 
 dependence
is reached for the longitudinal (e,e$'$p) results shown in
Fig.\ref{fig:ulmer}.  It is worth mentioning that the fact that 
the 1s strength has a high missing energy tail can be
attributed to correlation effects.  In order to reach the agreement of
Fig.\ref{fig:ulmer} a spectroscopic factor of 0.5 for the 1s1/2 state has
been adopted.  Despite the reasonable description of the longitudinal
structure function, the transverse strength remains underestimated
over the whole missing energy range.  Final-state interactions are a
possible explanation for this observation.  Eventhough differences in
the way FSI affects the different response functions have been noted, 
the FSI effects are unlikely to act in such a selective manner that
they produce a long tail of strength in the transverse response
function leaving the longitudinal one almost unaffected.

\begin{figure}
\centering
\epsfysize=12.cm
\epsffile{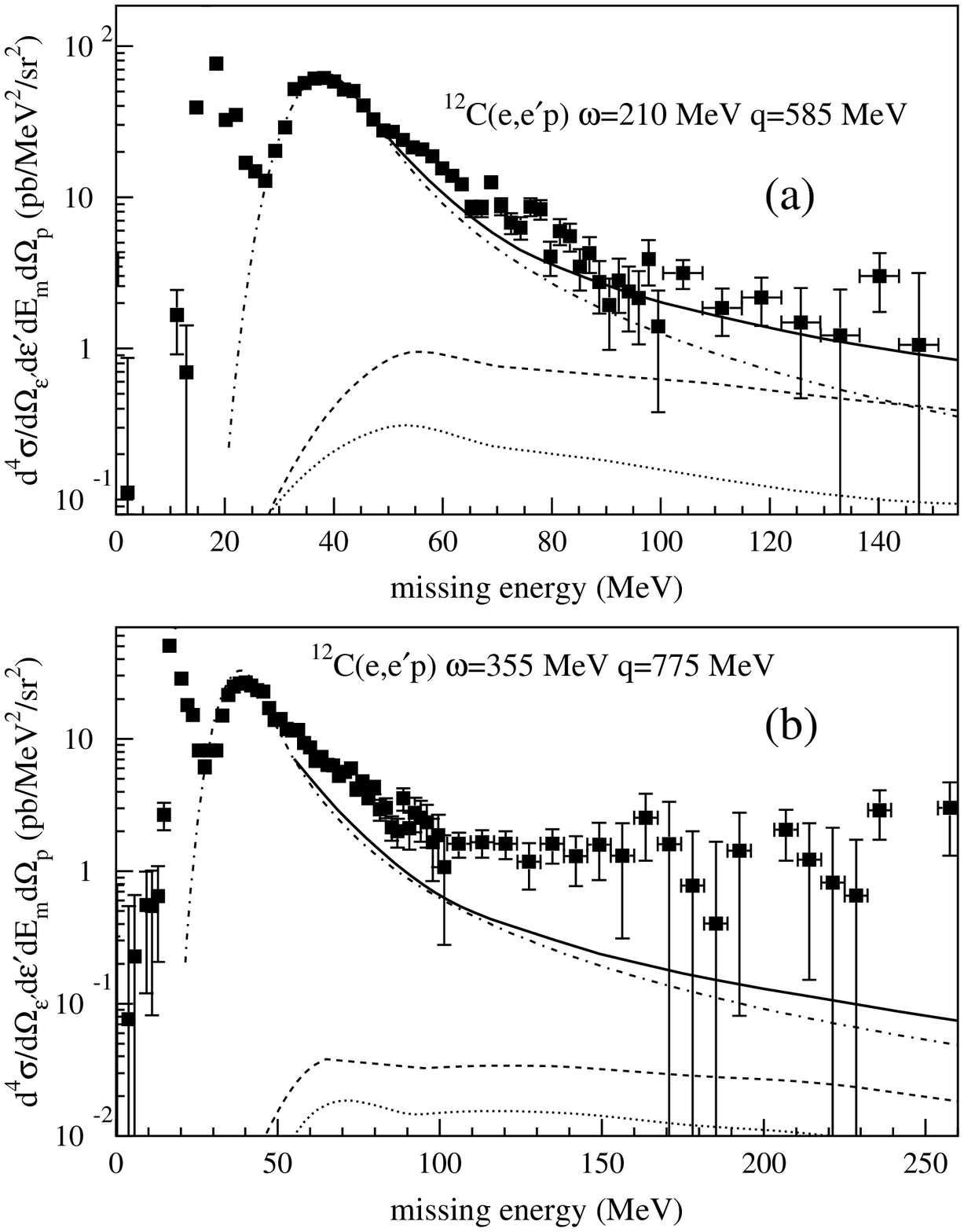}
\caption{Missing-energy spectra for $^{12}$C(e,e$'$p) at quasi-elastic
kinematics. (a) $\epsilon$=505.4~MeV, $\omega$=210~MeV and
q=585~MeV/c (b) $\epsilon$=698~MeV, $\omega$=355~MeV and q=775~MeV/c.
The dotted (dashed) line is the calculated strength from (e,e$'$pp)
((e,e$'$pn)).  The dot-dashed line the calculated strength from
exclusive one-proton ejection out of the $1s$ shell.  The solid is the
summed strength from all calculated contributions.  The data are from
Ref.~\protect \cite{Wei90}.} 
\label{fig:weinstein}
\end{figure}

In Fig.~\ref{fig:weinstein} we show a similar theoretical analysis for
the measurements of Ref.~\cite{Wei90} that were performed at higher
energy and momentum transfer.  The $^{12}$C(e,e$'$p) data of
Fig.~\ref{fig:weinstein} refer to quasi-elastic conditions and were
obtained in parallel kinematics ($\vec p _p \| \vec q$). 
As the data cover a larger
missing-energy range than those of Ref.~\cite{Ulm87} the approximation
(\ref{eq:appro}) is at stake.  The data show a clear bump in the
missing-energy region that is dominated by 
proton removal out of the 1s-shell.  Above
the s-shell region, the data exhibit a long tail.  Part of this strength can
be attributed to two-nucleon knockout.  For the largest momentum
transfer considered here, however, the calculated
two-nucleon knockout strength is just a small fraction of the measured
strength. At q=585~MeV/c a reasonable account of the data is achieved
when adding the calculated 2N knockout strength to the contribution
 related to knockout from the $1s$ shell.  In the
region 
just above the 1s bump
where the approximation (\ref{eq:appro}) is expected to be reasonable,
the calculations underestimate the data in both cases.   
It remains to be
investigated whether model calculations starting from realistic
spectral functions can explain the origin of this strength.

\section{Conclusion}
In this paper we have provided a framework in which
the effect of ground-state correlations on electroinduced one- and
two-nucleon knockout
cross sections could be quantified.  
We have gone beyond the standard
shell-model by considering short-range corrections to the nuclear wave
functions.  In the first-generation calculations presented here, this
has been achieved by a state-independent Jastrow ansatz.  The effect
of the short-range effects on the cross sections is estimated on the
basis of a cluster expansion.  It is shown that when considering
correlated wave functions, standard one-body photoabsorption in the
Impulse Approximation generates a whole chain of ``effective''
absorption mechanisms that are 2,3,..., A-body in nature.  These
multi-body effects can be related to the ground-state correlations and
compete with regular two-nucleon photoabsorption mechanisms, as e.g.
generated through meson-exchange and $\Delta _{33}$ creation. 

In the model outlined, a partial-wave expansion technique is adopted
for the description of the final-state interaction in A(e,e$'$NN)
processes. A direct knockout reaction model is adopted.  Special care
is taken to treat all spin degrees of freedom and anti-symmetrization
effects exactly.  The model is particularly suited to calculate
exclusive  
A(e,e$'$NN) angular cross sections for creation of the residual
nucleus in a specific state.  

A novel technique to calculate the 2N knockout contribution to the
semi-exclusive (e,e$'$N)  channel is outlined.  With the aid of Racah
algebra, the integration over the solid angle of the undetected 
escaping nucleon
could be done analytically.  This procedure
cuts severely on the amount of numerical integrations which have to be
performed. The procedure of calculating the semi-exclusive (e,e$'$p)
strength adopted here, has the advantage of being unfactorized in
nature. This allows  calculating the contribution from ground-state
correlations without excluding contributions from other sources, as
e.g.  these arising from photoabsorption on two-body currents.  Such
contributions are completely excluded when expressing the
semi-exclusive (e,e$'$p) cross section in terms of the one-body
spectral function.  

As far as the exclusive 2N knockout channel is concerned, the numerical
results presented here include both (e,e$'$pp) and (e,e$'$pn) angular
cross sections for the target nuclei $^{12}$C and  $^{16}$O.  The
shapes and magnitudes of the angular cross sections reflect the
complexity of the photoabsorption mechanism and the shell-model
structure of the target and residual nucleus.  The (e,e$'$pn) channel
is characterized by considerably larger cross sections than the
(e,e$'$pp) channel.  Electroinduced proton-neutron emission is
dominated by the mesonic and $\Delta$-isobar degrees of freedom, with
the central ground-state correlations playing only a marginal role.
Therefore, the proton-neutron knockout strength is 
predominantly transverse in nature even when longitudinal kinematics
is adopted.
Two-proton knockout, on the other hand, exhibits a selective
sensitivity to ground-state correlation effects.  A major source of
electroinduced two-proton knockout is, however, the  
$\Delta N \rightarrow pp$ mechanism.  It is therefore essential to have
a realistic description of these isobar degrees of freedom.  This
requires the input of knowledge obtained within the context of pion
and real-photon absorption on nuclei.

The conclusions drawn for the (e,e$'$NN) reaction have their
implications for the semi-exclusive (e,e$'$p) channel as correlation
effects will manifest themselves predominantly as two-nucleon
knockout.  The calculations predict this channel to be dominated by
proton-neutron knockout, which in its turn is strongly fed through
the meson-exchange and isobaric currents. Apart from the two-body
currents considered here, also multi-scattering effects can
contaminate the link between the semi-exclusive (e,e$'$p) data and the
spectral function.  This effect has been the object of several
investigations \cite{Tak89,Sick} with sometimes different conclusions.  
All this puts heavy constraints
on the applicability of the semi-exclusive (e,e$'$p) reaction to
gain empirical information 
about the single-particle spectral function.  In any case, the
strength generated by meson-exchange and $\Delta$ degrees of freedom
has to be carefully estimated and 
substracted from the (e,e$'$p) data before relating the measured
strength to ground-state correlation effects.
Better conditions to detect the correlation effects in (e,e$'$p)  are
predicted to occur at higher momentum transfer and specific bands in
the ($E_m,p_m)$ configuration space.  
   
{\bf Acknowledgement}\\
The authors are grateful to Dr. L.B. Weinstein for kindly providing the
data files of the MIT measurements and stimulating discussions. 
This work has been supported by the Fund for Scientific
Research-Flanders (FWO).
\newpage

\appendix
\section{Two-body matrix elements with short-range currents}
The purpose of this Appendix is to give the expressions for the two-body
matrix elements (Eqs. (\ref{mfcoul}) and (\ref{mfel}))
with the effective two-body currents that account for the
short-range effects in the initial wave function (Eqs.~(\ref{srccur})
and (\ref{srccha})).  As outlined
in Sect.~\ref{sec:src} these effective operators are composed of the
one-body operators of the 
impulse approximation and a Jastrow like correlation function that
corrects for short-range correlations in the initial-state wave function. 

The longitudinal strength attributed to SRC is contained in the 
two-body charge density of Eq.~(\ref{srccha}) and is determined by the
following effective Coulomb operator (\ref{opcoul}) : 
\begin{eqnarray}
M_{JM}^{coul}\left( \rho _{SRC} ^{[2]}(1,2) \right) & = &
- \left( j_J (q r_1) Y_{JM} (\Omega _1) g(r_{12}) e \frac {1 + \tau
_{z,1}} {2} \right.
\nonumber \\
& &  + \left.
j_J (q r_2) Y_{JM} (\Omega _2) g(r_{12}) e \frac {1 + \tau
_{z,2}} {2} \right)  
\end{eqnarray}
With the aid of the expansion (\ref{gexpan}) this leads to the
following two-body matrix elements : 
\begin{eqnarray}
& & <ab ;J_1 \|  M_{J}^{coul}(\rho ^{[2]} _{SRC}(1,2)) \|cd ; J_2 > =
\nonumber \\
& & -  
\sum_{l L} \sqrt{4 \pi} e \frac{\widehat{J_1} \widehat{L}
 \widehat{J_2}}{\widehat{l}} 
< l \; 0 \; L \;0 \mid J \; 0> \int dr_1 \int dr_2 g_l^c(r_1,r_2)
\nonumber \\
& & \times [\delta _{ac,\pi} \; 
X(j_a,j_b,J_1;j_c,j_d,J_2;L,l,J) D(a,c,L,r_1) D(b,d,l,r_2) j_J(qr_1) 
\nonumber \\ 
& &  + \delta _{bd,\pi} \;
X(j_a,j_b,J_1;j_c,j_d,J_2;l,L,J) D(b,d,L,r_2) D(a,c,l,r_1) j_J(qr_2)]\;,
\end{eqnarray}   
with
\begin{equation}
D(a,b,l,r)   \equiv    <l_a j_a \| Y_{l}(\Omega) \|l_b j_b>_r  \;.
\end{equation}
The radial transition density $<a\|\widehat{O}\|b>_r$ is defined such that 
it is related to the full matrix element through $<a\|\widehat{O}\|b>
= \int dr <a\|\widehat{O}\|b>_r$. 
In the above expressions $X$ denotes the 9j symbol in
the conventions of ref.~\cite{talmi} :
\begin{eqnarray}
 X(j_a,j_b,J_1;j_c,j_d,J_2;j_e,j_f,J) \equiv \left\{ \begin{array}{lll}
    j_a \; j_b \; J_1  \nonumber \\
    j_c \; j_d \; J_2  \nonumber \\
    j_e \; j_f \; J
    \end{array} \right\} \;.
\end{eqnarray}
Introducing the operator 
\begin{eqnarray}
O_{JM}^{\kappa}(q) & = & \sum_{M_1,M_2} 
\int d{\vec r}
 < J+\kappa \; M_1 \; 1 \; M_2 | J \; M > 
\nonumber \\
& & \times Y_{J+\kappa
M_1} (\Omega) j_{J+\kappa}(qr) J_{M_{2}}({\vec r})\;, 
\end{eqnarray}
the electric and magnetic transition operators (Eq.~\ref{opel}) can be
rewritten as :
\begin{eqnarray}
T_{JM}^{mag}(q) & = &  O_{JM}^{\kappa =0} \\
T_{JM}^{el}(q)  & = & \sum_{\kappa =\pm 1} 
\frac{i(-1)^{\delta_{\kappa,+1}}} {\widehat{J}} \sqrt{J+\delta_{\kappa,-1}}
O_{JM}^{\kappa}\;.
\end{eqnarray}
Accordingly, the matrix elements of the operator $O_{JM}^{\kappa}$
suffice to determine both the electric and the magnetic strength.  The
transverse strength attributed to SRC is determined by the effective
two-body current of Eq.~(\ref{srccur}).  The latter has two components
related to the convection and magnetization one-body
current of the impulse approximation.  
For the component related to
the magnetization current the effective $O _{JM} ^{\kappa}$ operator
reads~: 
\begin{eqnarray}
& & O_{J}^{\kappa} \left({\vec J}_{SRC} ^{[2],magn}(1,2)\right)  =  -
\int d {\vec r} \sum _ {M_1 M_2} < J+\kappa \; M_1 \; 1 \; M_2 \mid J M >
\nonumber \\
& &  \times \left\{  
      \frac {\mu_1 e} {2 M_N} \delta ({\vec r} - {\vec r}_1) g(r_{12})
      \left( \left(\vec{\nabla} \times \vec{\sigma} _1
\right)_{M_{2}} 
Y_{J+\kappa
      M_{1}}(\Omega) j_{J+\kappa}(qr) \right) \right. 
\nonumber \\
& &    + \left. \frac {\mu_2 e} {2 M_N} \delta ({\vec r} - {\vec r}_2) g(r_{12})
      \left( \left(\vec{\nabla} \times 
 \vec{\sigma} _2 \right)_{M_{2}} Y_{J+\kappa
      M_{1}}(\Omega) j_{J+\kappa}(qr) \right) \right\}
\end{eqnarray}                      
After some lenghty but straightforward manipulations, the transition
matrix element for this operator can be reduced to :
\begin{eqnarray}
& & <ab ;J_1 \|  O_{J}^{\kappa} \left({\vec J}_{SRC} ^{[2],magn}(1,2)
\right) \|cd ; J_2 > = \sqrt{6 \pi}
\frac{ieq}{M_N} \nonumber \\
& & \times 
 \sum_{\eta=\pm 1}
\sum_{l L J_4} 
\frac{\widehat{J_1} \widehat{L} \widehat{J_2} \widehat{J} \widehat{J_4}}
 {\widehat{l}}
< l \; 0 \; L \;0 \mid J+\kappa+\eta \; 0>
\nonumber \\
& & \times \sqrt{J+\kappa+\delta_{\eta,+1}}
\left\{ \matrix{ J & 1 & J+\kappa+\eta \nonumber \cr
                 L & l & J_4 \nonumber \cr } \right\}\;
\left\{ \matrix{ J+\kappa & J+\kappa+\eta & 1 \nonumber \cr
                 1 & 1 & J \nonumber \cr } \right\}                 
\nonumber \\
& & \times \int dr_1 \int dr_2 g_l^c(r_1,r_2)
\left\{ [\mu _p \delta _{ac,\pi} + \mu _n \delta
_{ac,\nu}] (-1)^{l+J+J_{4}} \right.
\nonumber \\
& & \times     X(j_a,j_b,J_1;j_c,j_d,J_2;J_4,l,J)
B(a,c,J+\kappa+\eta,L,J_4,qr_1) D(b,d,l,r_2) 
\nonumber \\ 
& & +  \left[ \mu _p \delta _{bd,\pi} + \mu _n \delta _{bd,\nu}
\right] X(j_a,j_b,J_1;j_c,j_d,J_2;l,J_4,J) \nonumber \\
& & \left. \times B(b,d,J+\kappa+\eta,L,J_4,qr_2)] D(a,c,l,r_1)
\right\} \;,
\end{eqnarray}
where we have introduced the radial transition density
\begin{eqnarray}
B(a,b,l,J_1,J_2,pr)  & \equiv &
<a \| j_l(pr) \left[Y_{J_{1}}(\Omega) \otimes
\vec{\sigma} \right]_{J_{2}} \| b >_{r} \nonumber \\ 
& =  &\widehat{j_a} \widehat{J_2} \widehat{j_b}
X(l_a,1/2,j_a;l_b,1/2,j_b;J_1,1,J_2)
\widehat{l_a} \widehat{J_1} \sqrt{\frac{3}{2 \pi}} \nonumber \\
 & & \times (-1)^{J_1}< l_a \; 0 \; J_1 \; 0 | l_b \; 0> r^2
\varphi _a (r) \varphi _b (r)j_l(pr) \nonumber 
\end{eqnarray}
The matrix element for the SRC contribution related to the convection
current can be derived in an analoguous manner and reads :
\begin{eqnarray}
& & <ab ;J_1 \|  O_{J}^{\kappa}\left({\vec J}_{SRC} ^{[2],conv}(1,2)
\right) \|cd ; J_2 > = 
\nonumber \\
& & \sqrt{\pi} \frac{e}{i M_N}
\sum_{l L J_4} 
\frac{\widehat{J_1}  \widehat{J_2} \widehat{J} \widehat{J_4}
\widehat{L}}{\widehat {l}} 
 < l \; 0 \; L \;0 \mid J+\kappa \; 0>
\nonumber \\
& & \times 
\left\{ \matrix{ J & 1 & J+\kappa \nonumber \cr
                 L & l & J_4 \nonumber \cr } \right\}\; (-1)^{L+J_4}
\int dr_1 \int dr_2 g_l^c(r_1,r_2)
\nonumber \\
& & \times \left\{ \delta _{ac,\pi} 
X(j_a,j_b,J_1;j_c,j_d,J_2;J_4,l,J) E(a,c,J+\kappa,L,J_4,qr_1) D(b,d,l,r_2) 
\right. \nonumber \\ 
& &  + \delta _{bd,\pi} (-1)^{l+J_4+J}
X(j_a,j_b,J_1;j_c,j_d,J_2;l,J_4,J) 
\nonumber \\ 
& & \left. \times E(b,d,J+\kappa,L,J_4,qr_2) D(a,c,l,r_1)
\right\} \;,
\end{eqnarray}
where we have introduced the radial transition density
$E(a,b,l,J_1,J_2,pr)$ 
\begin{equation}
E(a,b,l,J_1,J_2,pr)  \equiv 
<a \| j_l(pr) \left[ Y_{J_{1}}(\Omega) \otimes
(\vec{\nabla} - \vec{\nabla} ') \right] _{J_{2}} \| b >_{r}  \; .
\label{eq:vgcon}
\end{equation}
The notation $\vec{\nabla} '$ refers to a gradient operator acting to
the left.  The terms involving the derivates of the central
correlation functions have been neglected in the above expression.

\newpage

\begin{thebibliography}{99}
\bibitem{mahan}
C. Mahaux and R. Sartor, Adv. Nucl. Phys. {\bf 20} (1991) 1.
\bibitem{pieper}
S.C. Pieper, R.B. Wiringa and V.R. Pandharipande, Phys. Rev. {\bf C46}
(1992) 1741.
\bibitem{co}
G. Co$'$, A. Fabrocini, S. Fantoni and I.E. Lagaris, Nucl. Phys. {\bf
A549} (1992) 439.
\bibitem{della}
F. Dellagiacoma, G. Orlandini and M. Traini, Nucl. Phys. {\bf A393}
(1983) 95.
\bibitem{gua96}
R. Guardiola, P.I. Moliner, J. Navarro, R.F. Bishop, A. Puente and
Niels R. Walet, Nucl. Phys. {\bf A609} (1996) 218.
\bibitem{wirz89}
A. Wirzba, H. Toki, E.R. Siciliano, Mikkel B. Johnson and R. Gilman,
Phys. Rev. {\bf C40} (1989) 2745.
\bibitem{johnson}
Mikkel B. Johnson, E.R. Siciliano and H. Sarafian, Phys. Lett. {\bf
B243} (1990) 18.
\bibitem{ciofi96}
C. Ciofi degli Atti and S. Simula, Phys. Rev. C {\bf 53} (1996) 1689.
\bibitem{mutter}
H. M\"{u}tter and W.H. Dickhoff, Phys. Rev. C {\bf 49} (1994) R17.
\bibitem{wei94}
L. Weinstein and W. Bertozzi, Proc. of the 6th Workshop on
Perspectives in Nuclear Physics at Intermediate Energies,
(ICTP-Trieste, Italy, May 3-7, 1993), Eds. S. Boffi, C. Ciofi degli
Atti and M. Giannini, World Scientific (1994), 362.
\bibitem{Wei89}
L.B. Weinstein and W. Bertozzi, {\em in} Proc. of the Fourth Workshop on
Perspectives in Nuclear Physics at Intermediate Energies (Trieste,
1988), Eds. S. Boffi, C. Ciofi degli Atti and M. Giannini, (World
Scientific, Singapore, 1989).
\bibitem{vemec}
V. Van der Sluys, J. Ryckebusch and M. Waroquier, Phys. Rev. C {\bf
49} (1994) 2695.
\bibitem{Got58}
Kurt Gottfried, Nucl. Phys. {\bf 5} (1958) 557.
\bibitem{cross}
G.E. Cross {\it et al.}, Nucl. Phys. {\bf A593} (1995) 463.
\bibitem{grab96}
P. Grabmayr {\em et al.}, Phys. Lett. {\bf B370} (1996) 17.
\bibitem{lamp96}
Th. Lamparter {\em et al.}, Z. Phys. {\bf A355} (1996) 1.
\bibitem{harty96}
P.D. Harty {\em et al.}, Phys. Lett. {\bf B380} (1996) 247.
\bibitem{zondervan}
A. Zondervan {\em et al.}, Nucl. Phys. {\bf A587} (1995) 697.
\bibitem{leon}
L.J.H.M. Kester {\em et al.}, Phys. Rev. Lett. {\bf 74} (1995) 1712.
\bibitem{jannpa}
J. Ryckebusch, M. Vanderhaeghen, L. Machenil and M. Waroquier, Nucl.
Phys. {\bf A568} (1994) 828.
\bibitem{Waro}
M. Waroquier, J. Ryckebusch, J. Moreau, K. Heyde, N. Blasi,
S.Y. van der Werf and G. Wenes, Phys. Rep. {\bf 148} (1987) 249. 
\bibitem{geurts}
W.J.W. Geurts, K. Allaart, W.H. Dickhoff, H. M\"uther,
Phys. Rev. C {\bf 54} (1996) 1144.
\bibitem{ryc94}
J. Ryckebusch, L. Machenil, M. Vanderhaeghen, V. Van der Sluys and M.
Waroquier, Phys. Rev. {\bf C49} (1994) 2704.
\bibitem{mach}
R. Machleidt {\em in} Relativistic Dynamics and Quark-Nuclear Physics,
Eds. M.B. Johnson and A. Picklesimer, (John Wiley and Sons, New York,
1986), 71.
\bibitem{marc}
M. Vanderhaeghen, L.Machenil, J. Ryckebusch and M. Waroquier, Nucl.
Phys. {\bf A580} (1994) 551.
\bibitem{Oset}
E. Oset, H. Toki and W. Weise, Phys. Rep. {\bf 83} (1982) 281.
\bibitem{dekker} 
M.J. Dekker, P.J. Brussaard and J.A. Tjon, Phys. Rev.
C {\bf 49} (1994) 2650.
\bibitem{osterfeld}
B. K\"{o}rfgen, P. Oltmanns, F. Osterfeld and T. Udagawa, Phys. Rev. C,
(1997) in press.
\bibitem{bianchi}
N. Bianchi {\em et al.}, Phys. Rev. C {\bf 54} (1996) 1688.
\bibitem{koch}
J.K. Koch and N. Ohtsuka, Nucl. Phys. {\bf A435} (1985) 765 and J.
Koch {\em in} Moderns Topics in Electron Scattering, Eds. B. Frois and
I. Sick, (World Scientific, Singapore) (1991) 28.
\bibitem{oconnell}
J.S. O'Connell {\em et al.}, Phys. Rev. C {\bf 35} (1987) 1063.
\bibitem{chen}
C.R. Chen and T.-S.H. Lee, Phys. Rev. C {\bf 38} (1988) 2187.
\bibitem{douglas}
I.J.D. MacGregor {\em et al.}, to be published.
\bibitem{thomas} 
T. Wilbois, P. Wilhelm, and H. Arenh\"ovel, Phys. Rev. C {\bf 54}
(1996) 3311. 
\bibitem{reply}
J. Ryckebusch, L. Machenil, M. Vanderhaeghen, V. Van der Sluys, and M.
Waroquier, Phys. Rev. C {\bf 54} (1996) 3313. 
\bibitem{riska83}
D.O. Riska, in Prog. Part. and Nucl. Physics, Vol. 11 (1982) 199.
\bibitem{gaudin}
M. Gaudin, J. Gillespie and G. Ripka, Nucl. Phys. {\bf A176} (1971) 237.
\bibitem{clark}
J.W. Clark, {\em in} The many-body problem : Jastrow correlations
versus Brueckner Theory, R. Guardiola and J. Ros eds., 
Lecture Notes in Physics {\bf 138} (Springer Verlag, Berlin, 1981) 184.
\bibitem{fantoni}
S. Fantoni and V.R. Pandharipande, Nucl. Phys. {\bf A473} (1987) 234.
\bibitem{fabro}
Adelchi Fabrocini, Phys. Lett. {\bf B322} (1994) 171.
\bibitem{vijay}
V.R. Pandharipande and R.B. Wiringa, Rev. Mod. Phys. {\bf 51} (1979) 821.
\bibitem{hodgson}
A.N. Antonov, P.E. Hodgson and I. Zh. Petkov, Nucleon Momentum and
Density Distributions in Nuclei (Clarendon Press, Oxford, 1988).
\bibitem{vanorden}
J.W. Van Orden and T.W. Donnelly, Ann. of Physics {\bf 131} (1981) 451.
\bibitem{co2}
G. Co$'$, A. Fabrocini and S. Fantoni, Nucl. Phys. {\bf A568} (1994) 73.
\bibitem{benhar}
O.Benhar, A. Fabrocini and S. Fantoni, in  {\em Modern Topics in Electron
Scattering}, eds. B. Frois and I. Sick (World Scientific, Singapore, 1991),
460.
\bibitem{omy}
T. Ohmura, M. Morita and M. Yamada, Prog. Theor. Phys. {\bf 15} (1956)
222.
\bibitem{janfac}
J. Ryckebusch, Phys. Lett. {\bf B383} (1996) 1.
\bibitem{sick94}
I. Sick, S. Fantoni, A. Fabrocini and O. Benhar, Phys. Lett. {\bf
B323} (1994) 323.
\bibitem{benh94}
O. Benhar, A. Fabrocini, S. Fantoni and I. Sick, Nucl. Phys. {\bf
A579} (1994) 493.
\bibitem{janold}
Jan Ryckebusch, Marc Vanderhaeghen, Kris Heyde and Michel Waroquier,
Phys. Lett. {\bf B350} (1995) 1.
\bibitem{anghi}
M. Anghinolfi {\em et al.}, Nucl. Phys. {\bf A457} (1986) 645.
\bibitem{leonplb}
L.J.H.M. Kester {\em et al.}, Phys. Lett. {\bf B344} (1995) 79.
\bibitem{Wei90}
L.B. Weinstein {\it et al.}, Phys. Rev. Lett. {\bf 64} (1990) 1646.
\bibitem{Ulm87}
P.E. Ulmer {\it et al.}, Phys. Rev. Lett. {\bf 59} (1987) 2259.
\bibitem{highpm}
V. Van der Sluys, J. Ryckebusch and M. Waroquier, Phys. Rev. {\bf C54}
(1996) 1322.
\bibitem{Jeu83}
J.P. Jeukenne and C. Mahaux, Nucl. Phys {\bf A394} (1983) 445.
\bibitem{Tak89}
T. Takaki, Phys. Rev. {\bf C39} (1989) 359.
\bibitem{Sick}
I. Sick, preprint University of Basel.
\bibitem{talmi}
I. Talmi, Simple Models of Complex Nuclei : The Shell Model and
Interacting Boson Model, (Harwood Academic Publishers, Chur, 1993).
\end{thebibliography}
\end{document}